\documentclass[a4paper,
               onecolumn,
               10pt,
               accepted=2021-12-02]{quantumarticle}
\pdfoutput=1
\usepackage[utf8]{inputenc}
\usepackage[english]{babel}
\usepackage[T1]{fontenc}

\usepackage{tabularx}
\usepackage{amsmath}
\usepackage{amssymb}
\usepackage{mathtools}
\usepackage{braket}
\usepackage{graphicx}
\graphicspath{{figures/}}
\usepackage{calc}
\usepackage[dvipsnames]{xcolor}

\usepackage[numbers,sort&compress]{natbib}

\usepackage[colorlinks,urlcolor=blue,linkcolor=blue,citecolor=blue]{hyperref}

\usepackage{mathbbol}
\usepackage{dsfont}
\usepackage{siunitx}
\usepackage{calc}
\usepackage{comment}
\usepackage[normalem]{ulem}

\usepackage{array}
\newcolumntype{L}{>{$}l<{$}}
\newcolumntype{C}{>{$}c<{$}}

\newcommand\blfootnote[1]{%
\begingroup
\renewcommand\thefootnote{}\footnote{#1}%
\addtocounter{footnote}{-1}%
\endgroup
}

\newcommand{\nwsearrow}{\mathrel{\text{$\nwarrow$\llap{$\searrow$}}}}

\begin{document}

\title{Fundamental thresholds of realistic quantum error correction circuits from classical spin models}

\author{Davide Vodola$^{*,}$}
\affiliation{Dipartimento di Fisica e Astronomia ``Augusto Righi'' dell'Universit\`a di Bologna, I-40127 Bologna, Italy}
\affiliation{INFN, Sezione di Bologna, I-40127 Bologna, Italy}

\author{Manuel Rispler$^{*,}$}
\blfootnote{$^{*}$These two authors contributed equally to the present work.}
\affiliation{QuTech, Delft University of Technology, Lorentzweg 1, 2628 CJ Delft, The Netherlands}

\author{Seyong Kim}
\affiliation{Department of Physics, Sejong University, 05006 Seoul, Republic of Korea}

\author{Markus M\"uller}
\affiliation{Institute for Theoretical Nanoelectronics (PGI-2), Forschungszentrum Jülich, 52428 Jülich, Germany}
\affiliation{Institute for Quantum Information, RWTH Aachen University, 52056 Aachen, Germany}

\maketitle

\begin{abstract}
Mapping the decoding of quantum error correcting (QEC) codes to classical disordered statistical mechanics models allows one to determine critical error thresholds of QEC codes under phenomenological noise models. Here, we extend this mapping to admit realistic, multi-parameter noise models of faulty QEC circuits, derive the associated strongly correlated classical spin models, and illustrate this approach for a quantum repetition code with faulty stabilizer readout circuits. We use Monte-Carlo simulations to study the resulting phase diagram and benchmark our results against a minimum-weight perfect matching decoder. The presented method provides an avenue to assess fundamental thresholds of QEC circuits, independent of specific decoding strategies, and can thereby help guiding the development of near-term QEC hardware.
\end{abstract}

The interdisciplinary endeavour to develop large-scale quantum computers has revealed unexpected and strong connections across fields, in particular between quantum information theory and statistical physics. Exploring these links has been fruitful in both directions, e.g.~by efficient quantum algorithms to estimate partition functions of classical spin systems \cite{Lidar1997,Somma2007,Nest2007,Geraci2008,Cuevas2009,Cuevas2011,Xu2011,Cuevas2016,Zarei2018}, by providing effective descriptions for entanglement spreading in random circuits~\cite{Li2020,Nahum2018,Zhou2019} or in the context of quantum error correction (QEC). For the latter, it has been shown that the decoding of leading QEC codes \cite{Terhal2015,Lidar2013} such as the topological surface \cite{Kitaev2003,Dennis2002} and color codes \cite{Bombin2006,Bombin2007} can be mapped onto disordered classical statistical mechanics models. Locating the phase transitions between ordered and disordered phases in these models reveals the parameter regimes for which QEC succeeds or fails, respectively \cite{Dennis2002}. So far, however, these mappings have been largely limited to QEC codes with simple phenomenological noise models: for instance, uncorrelated \cite{Dennis2002, Katzgraber2009,Katzgraber2010, Bombin2012, Andrist2016} or weakly correlated bit and/or phase flip errors \cite{Chubb2019}, phenomenological measurement errors during syndrome readout \cite{Ohno2004,Andrist2011}, qubit loss and leakage \cite{Ralph2005,Stace2009,Stace2010,Vodola2018,Amaro2020} have been considered, with recent extensions also to bosonic QEC codes \cite{Vuillot-PhysRevA.99.032344}. Realistic modeling of experimental quantum hardware \cite{Corcoles2015,Saffman_2016,Negnevitsky2018,Andersen2019, Bruzewicz2019,Kjaergaard_2020,Andersen2020,Pezzagna,Chatterjee2021,Egan2021}, however, requires multi-parameter noise model descriptions of failure processes of the underlying quantum circuit components.

In this work, we extend the QEC-statistical mechanics model mapping and apply it to realistic quantum circuit-noise scenarios. For Clifford measurement circuits, correlations arising from circuit noise can be efficiently related to effective phenomenological models as direct input for (possibly sub-optimal) decoders \cite{Chubb2019,Pryadko2020}. Here, we use a similar approach but perform a systematic error propagation analysis and identify emerging effective noise processes in realistic quantum circuits. Instead of direct decoding, we derive the associated disordered classical spin model with correlated quenched disorder, determine its phase diagram and thereby obtain maximum threshold values agnostic to specific decoding strategies.
This method allows one to assess the maximum potential of realistic state-of-the-art QEC architectures. 
This is particularly pressing for QEC codes when (near-)optimal decoding is computationally hard \cite{Iyer2015}, when optimal decoders are unknown or strongly dependent on the chosen QEC circuitry and noise model details~\cite{Stephens2014}, e.g.~in measurement circuits for high-weight stabilizers in color codes~\cite{Wang2010b,Landahl2011,Beverland2021} or recently developed flag-qubit based QEC~\cite{chaoReichardt,Chamberland_2018_q,Chamberland_2020_2,Chamberland_2020_1}.

We illustrate the mapping of circuit-level QEC codes to statistical mechanics models for a 1D quantum repetition code with faulty circuitry described by a multi-parameter microscopic noise model accounting for a number of single- and two-qubit gate noise processes. Whereas this code does not enable correction of arbitrary errors, current experimental efforts focus on achieving repetitive QEC with this code in the regime of beneficial error suppression for increasing code sizes~\cite{Schindler2011, Waldherr_2014, Kelly2015, Chen2021}.
We show that in contrast to oversimplified phenomenological noise models, new effective correlated noise processes arise, which we systematically derive and quantitatively relate to error rates of the underlying microscopic noise model. 

We anticipate a robust performance gap between efficient, though sub-optimal minimum-weight perfect matching (MWPM) decoding and the fundamental upper limit as established via the phase diagram of the statistical physics model. This demonstrates the viability of this complete mapping approach to identify maximal performance of QEC circuitry.

\begin{figure}
    \centering
        \includegraphics[width=\textwidth]{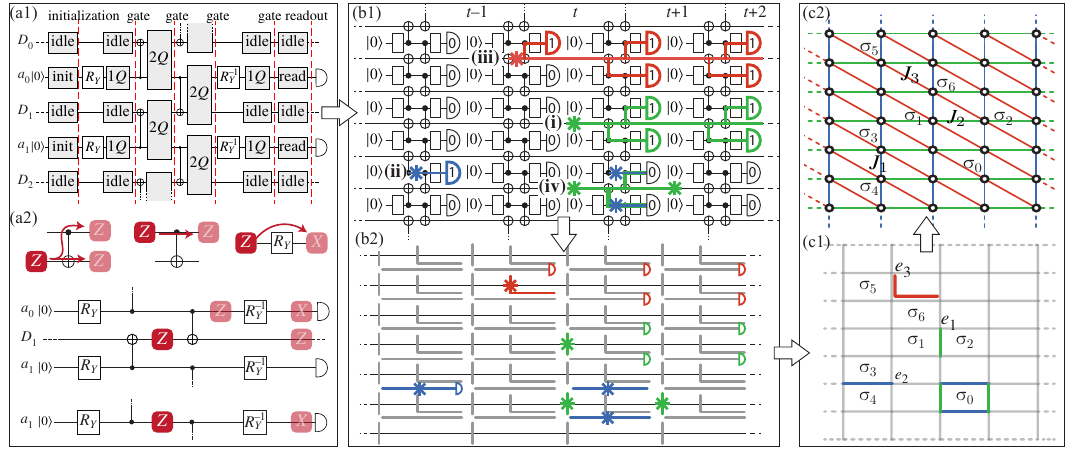}
    \caption{(a1) One cycle of the phase-flip repetition code with circuit-level noise, shown for three data qubits $D_i$ and two ancilla qubits $a_i$. White boxes $R_Y$ ($R_Y^{-1}$) represent single-qubit $Y$-rotations of $\pi/2$ ($-\pi/2$). Gray boxes (idle, init, 1$Q$, 2$Q$ and read) are depolarizing channels describing faults during idling, state preparation, single-qubit, CNOT gates and measurements, respectively. (a2) (Top row) Examples of error propagation through CNOT and $R_Y$ gates: $Z$ errors on target qubits propagate to the control qubits; $Z$ errors do not propagate from control to target qubits; $Z$ errors are transformed into $X$ errors through $R_Y$. (Center row) A $Z$ error between two CNOTs on $D_1$ propagates also as a measurement error to the ancilla $a_0$. (Bottom row) A $Z$ error between two $R_Y$ on the ancilla results in a measurement error.  (b1) Four measurement cycles showing error propagation and the affected ancilla qubits. (i) A data-qubit error (green asterisk) propagates along the green path and affects two ancilla qubits at every subsequent measurement. (ii) A measurement error (blue asterisk) affecting a single ancilla. (iii) An error between the CNOT gates of a data qubit (red asterisk) propagates along the red path and affects one ancilla at step $t$ and two ancilla qubits in subsequent measurements. (iv) A space-time equivalence given by two qubit errors and two measurement errors, which does not trigger any ancilla. (b2) Error graph generated by the physical errors in panel (b1). Positions where qubit, measurement and correlated data phase-flip and measurement errors can happen are represented by vertical, horizontal and L-shaped edges,  respectively. Colored semicircles represent ancilla qubits triggered by an error event. From this graph we derive the fundamental error events determining the couplings of the statistical mechanics Hamiltonian (panels (c1)-(c2)) and the syndrome volume. (c1) Fundamental effective errors $e_1$, $e_2$, $e_3$ generated by the error processes (i)-(iii) and equivalences $\sigma_\ell$. (c2) Lattice of equivalences $\sigma_\ell$ showing the couplings $J_1$ (vertical blue links), $J_2$ (horizontal green links), $J_3$ (diagonal red links) of the Hamiltonian of Eq.~\eqref{eqn_stat-mech_hamiltonian}.}
    \label{fig_panels}
\end{figure}

\section{Noisy quantum error correcting circuits}

Here we consider the phase-flip $n$-qubit repetition code \cite{Devitt_2013,nielsen-book} from the class of stabilizer QEC codes~\cite{Gottesman1997}. The $\pm1$ eigenvalues of the $n-1$ stabilizers $S_i = X_i\otimes X_{i+1}$ (with $X_i$ the Pauli X matrix on qubit~$i$) form the error syndrome, with error states having at least one non-trivial syndrome. In order to diagnose potential errors, the stabilizers $S_i$ are measured by coupling them to $n-1$ \emph{ancilla} qubits. 
Phase-flip errors $Z_i$ are detected, unless they happen on all qubits, which would be indistinguishable from a logical phase-flip $Z_L = \otimes_{i=0}^{n-1} Z_i$. Stabilizer measurements on real hardware are noisy as well and thus need to be repeated periodically. Analysing the discrete difference in time between successive measurement rounds results in another repetition code in the time domain, which yields a (1+1)-dimensional space-time \textit{syndrome volume}. The code-capacity critical error threshold, below which successful QEC is feasible, is $p_c=0.5$, when the only error source are data qubit errors. The phenomenological noise threshold, where errors are injected  with a rate $p$ per round on both ancilla and data qubits is $p_c \approx 0.11$~\cite{Honecker2001}. This is identical to 
the threshold of the toric code with perfect measurements that exactly maps onto the repetition code with phenomenological noise, essentially trading one spatial dimension for the time dimension introduced by repeating stabilizer measurements in time. 

\subsection{Microscopic circuit noise}

In contrast to the phenomenological noise models,
which ignore the physical reality of the circuit the QEC code is running on, in this work we analyze a multi-parameter circuit level noise model that treats every component of the circuit as faulty, as illustrated in Fig.~\ref{fig_panels}(a1).
Here, every operation is followed by a depolarizing noise channel (grey boxes in Fig.~\ref{fig_panels}(a1)). For single-qubit operations, we inject $P_i$ ($P_i\in {X,Y,Z}$) with probability $\lambda/3$, where $\lambda$ can represent state preparation ($p_\text{sp}$), idling ($p_\text{id}$), single-qubit gates~($p_1$) and measurements~($p_\text{m}$)~\cite{Raussendorf2007, Stephens2014}. Likewise, after a CNOT we inject a non-trivial one- and two-qubit Pauli operator $P_i\otimes P_j$ with probability $p_2/15$. Errors propagate according to the rules described in Fig.~\ref{fig_panels}(a2). For the special case $p_\text{sp}=p_\text{id}=p_1=p_\text{m}=p_2 = \lambda$ the threshold was found to lie at $\lambda\approx 0.033$ for maximum likelihood decoding~\cite{Rispler2020}.

\subsection{Effective noise processes}
Any microscopic circuit-level error configuration of the multi-parameter noise model will lead to one of three effective error processes or combinations thereof, called error chains $E$, with associated error probabilities: (i) a single data qubit phase flip error with probability $p$, (ii) a measurement error with probability $q$, and (iii) a correlated error flipping a data-qubit and simultaneously one of the adjacent measurement outcomes with probability $r$ (examples shown in Fig.~\ref{fig_panels}(b1)). Through propagation of Pauli error generators and factorizing the effective error channel(s) we can express an arbitrary Pauli circuit noise model in terms of the three effective error rates $p$, $q$ and $r$ (see Appendix~\ref{sec_derivation_circuit_model} for details):
\begin{align}
    \begin{split}
    &p= \frac{1}{2} \left[1- \left(1-\frac{16p_2}{15}\right)\left(1-\frac{4p_\text{id}}{3}\right)^4 \right],\\
    &q= \frac{1}{2} \left[1- \left(1-\frac{16p_2}{15}\right)\left(1-\frac{4p_1}{3}\right)^2\left(1-\frac{4p_\text{sp}}{3}\right)\left(1-\frac{4p_\text{m}}{3}\right) \right],\\
    &r=\frac{8}{15}p_2.
    \end{split}
      \label{eq:effective_error_rates}
\end{align}
Let us emphasize that this correspondence is exact: given an arbitrary set of circuit noise parameters, the above relationship expresses this set of noise parameters in terms of $p,q$ and $r$ exactly to all orders. We gather the error events from the microscopic processes (Fig.~\ref{fig_panels}(b1)) in the error graph of Fig.~\ref{fig_panels}(b2): here vertical, horizontal and L-shaped lines indicate positions where a data-qubit (i), a measurement (ii) or a correlated error (iii) event has occurred, while colored semicircles indicate the stabilizers triggered by these, respectively. The error graph is used to construct the types of interactions appearing in the classical statistical mechanics Hamiltonian (Fig.~\ref{fig_panels}(c1)) and to construct the syndrome volume as input for the MWPM decoder.

From the error graph, we can visualize the error processes by introducing the lattice in Fig.~\ref{fig_panels}(c1) where vertical links represent data qubits and horizontal links are measurement steps. The three error processes (i)-(iii) define the fundamental errors events $e_1, e_2, e_3$ shown in Fig.~\ref{fig_panels}(c1): 
\begin{enumerate}
\item the error $e_1$ is generated by a data qubit error (i) or a combination of a measurement error (ii) and a correlated error (iii) with probability 
\begin{equation}
\Pr(e_1) = p(1-q)(1-r) + (1-p)q r;
\end{equation}
\item the error $e_2$ is generated by a measurement error (ii) or a combination of a data qubit error (i) and a correlated error (iii) with probability 
\begin{equation}
\Pr(e_2) = q (1-p)(1-r) + r p (1-q);
\end{equation}
\item the error $e_3$ is generated by a correlated error (iii) or a combination of a data qubit error (i) and a measurement error (ii) with probability 
\begin{equation}
\Pr(e_3) = r (1-p)(1-q) + p q (1-r).
\end{equation}
\end{enumerate}

These fundamental errors form error chains $E$ that can be parametrized by introducing binary variables  $h_\ell,v_\ell\in\{\pm 1\}$ for each link $\ell$ of the lattice. $h_\ell$ ($v_\ell$) are set to $-1$ if the horizontal (vertical) link $\ell$ belongs to $E$.

\section{Statistical mechanics model}

Mapping an error model into a statistical mechanics Hamiltonian allows us to compute the probability $\text{Pr}(\bar{E})$ of the class $\bar{E}$ formed of all the errors that 
differ from a reference (candidate) error chain~$E$ by a so-called \textit{space-time equivalence}~\cite{Dennis2002, Chubb2019}: A space-time equivalence $\hat{\sigma}$ (simply called equivalence in the following) is a trivial sequence of errors not detected by the stabilizers and thus producing the same space-time syndrome. For the phase flip code, equivalences are given by the action of a first phase flip on a qubit $D_i$ at a measurement step~$t$, followed by two measurement errors on the ancilla qubits of the stabilizers $S_{i}$ adjacent to $D_i$ and a second phase flip error on $D_i$ before the measurement step $t+2$ (see (iv) in Fig.~\ref{fig_panels}(b1)). We associate a classical Ising spin $\sigma$ to each equivalence (e.g.~$\sigma_0$ in Fig.~\ref{fig_panels}(c1)) such that a spin configuration with $\sigma = -1$ effectively describes the error $E' = E \hat{\sigma}$ that differs from $E$ by the equivalence $\hat{\sigma}$. Thus, once we fix $E$, sampling over all possible spin configurations is equivalent to sampling over all possible errors $E'$ in the class $\bar{E}$, and the probability of the class $\bar{E}$ is $\text{Pr}(\bar{E}) = \sum_{\{\hat{\sigma}\}} \text{Pr}({E} \hat{\sigma})$. The limiting behavior of the complementary probabilities $\text{Pr}(\bar{E}) $ and $ \text{Pr}(\overline{E Z_L})$ with increasing system size allows us to distinguish two regimes: below threshold, either of the probabilities converges to unity and hence applying a correction from the same error class will undo the error with probability approaching unity, above threshold all probabilities remain asymptotically finite, i.e.~there is no decoding choice that removes the error with high probability~\cite{Bombin2012}. 

\begin{figure}
    \centering
    \includegraphics[width=0.6\textwidth]{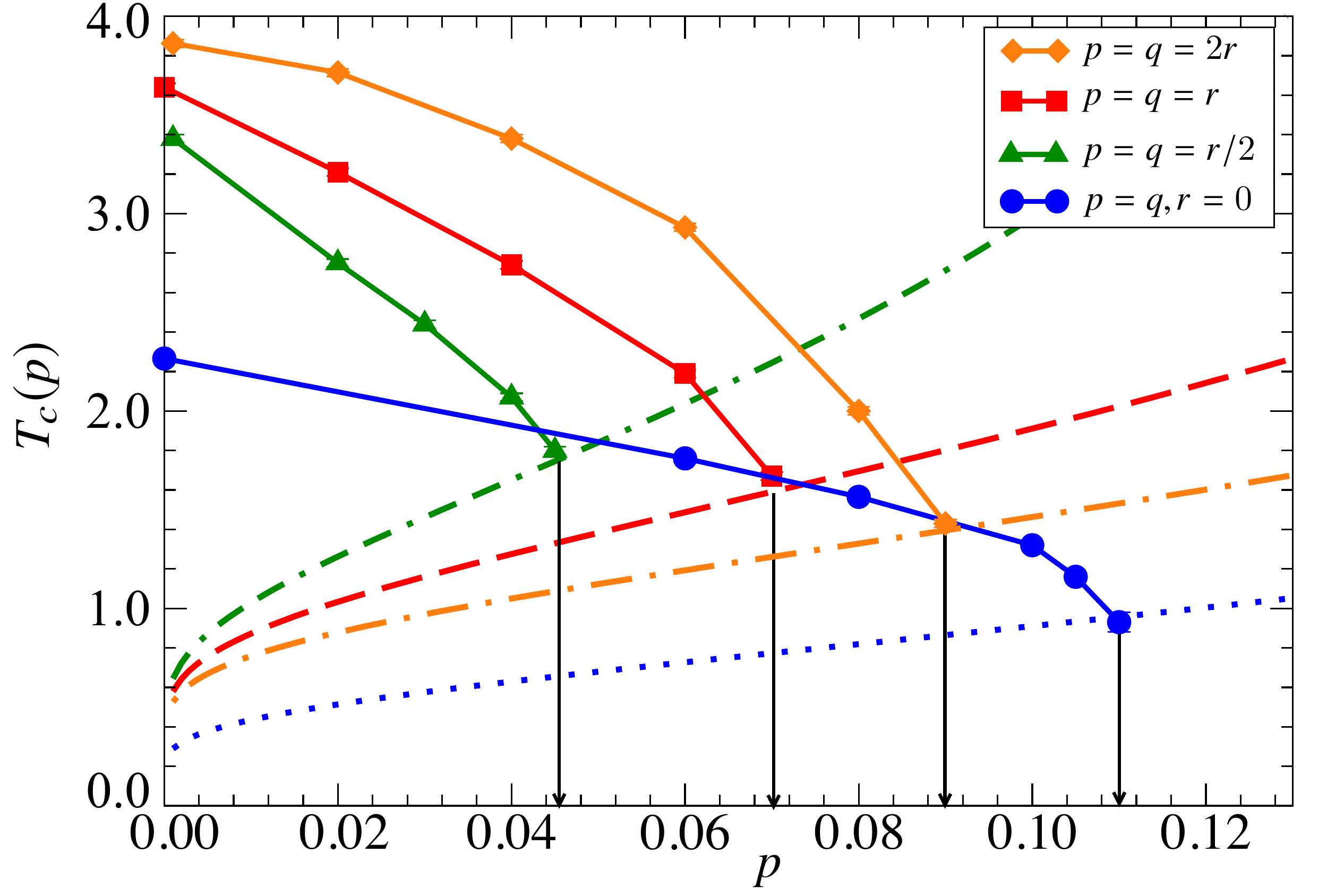}
    \caption{Phase diagram of the random bond Ising model for (I) $p = q = 2r$ (orange diamonds), (II) $p = q = r$ (red squares), (III) $p = q = r/2$ (green triangles), (IV) $p = q, r = 0$ (blue circles). The points represent the critical temperature $T_c$ for the given disorder configuration. Solid lines are a guide to the eye, and dashed lines are the Nishimori lines for each case. The vertical black lines indicate the locations of the largest $p$ for which $T_c$ can be determined. Error bars are smaller than the symbol size.}
    \label{fig:phasediagram}
\end{figure}

Given $E$ which is described by $h_\ell \in \{\pm 1\}$ and $v_\ell \in \{\pm 1\}$ and the set of equivalences $\{\sigma\}$, the probability of the error $E'$ can be written as $\text{Pr}(E') = \exp[-\beta H_E(\sigma)]$ where 
\begin{equation}\label{eqn_stat-mech_hamiltonian}
H_E(\sigma)  =  -\sum_{\ell \updownarrow\ell'} J_1 h_\ell  \sigma_\ell \sigma_{\ell'} - \sum_{\ell \leftrightarrow \ell'} J_2 v_\ell  \sigma_\ell \sigma_{\ell'} - \sum_{\ell \nwsearrow \ell' } J_3 h_\ell v_\ell  \sigma_\ell \sigma_{\ell'}
\end{equation}
is an Ising Hamiltonian with correlated quenched disorder. The couplings $J_1$, $J_2$, $J_3$ and the inverse temperature $\beta$ are fixed from the error model describing $E$ by the so-called Nishimori conditions (see Eq.\eqref{supp_Nishimori} in Appendix~\ref{sec_derivation_stat_mech}). Ferromagnetic or antiferromagnetic interactions are given by the signs of $v_\ell$ and $h_\ell$ fixed by the error $E$. In the first sum of the r.h.s.~of Eq.~\eqref{eqn_stat-mech_hamiltonian}, $\sigma_\ell$ and $\sigma_{\ell'}$ are the spins horizontally adjacent to the link $\ell$ (e.g. spins $\sigma_1$ and $\sigma_2$ in Fig.~\ref{fig_panels}(c1)); in the second sum, $\sigma_\ell$ and $\sigma_{\ell'}$ are the spins vertically adjacent to the link $\ell$ (e.g. spins $\sigma_3$ and $\sigma_4$ in Fig.~\ref{fig_panels}(c1)); in the third sum, $\sigma_\ell$ and $\sigma_{\ell'}$ are the two spins  adjacent to the two arms of the L-shaped error $e_3$ (e.g. spins $\sigma_5$ and $\sigma_6$ in Fig.~\ref{fig_panels}(c1)). A similar Hamiltonian has been analyzed for the surface code with spatially correlated phenomenological noise \cite{Chubb2019}, whereas here Hamiltonian \eqref{eqn_stat-mech_hamiltonian} arises from multi-parameter circuit noise.

\section{Numerical results}
\subsection{Monte Carlo simulations}

Mapping the QEC code to Eq.~\eqref{eqn_stat-mech_hamiltonian} allows us to access the critical thresholds of the code from thermodynamic properties of $H_E(\sigma)$~\cite{Dennis2002}. The error threshold separating the correctable from non-correctable parameter regime of the code corresponds to the phase transition between ordered and disordered phases of $H_E(\sigma)$. We determine this transition from Monte Carlo (MC) simulations and a scaling analysis of the two-body spin correlation length for various disorder strengths (see Appendix~\ref{sec_monte_carlo_details} for details). To this end, we transform the lattice in Fig.~\ref{fig_panels}(c1) into the triangular one of Fig.~\ref{fig_panels}(c2) where classical spins (white circles) reside on vertices and couplings $J_1, J_2, J_3$ are represented by vertical (blue), horizontal (green) and diagonal (red) links, respectively.

Figure~\ref{fig:phasediagram} shows the resulting phase diagram for (I) $p=q=2r$, (II) $p=q=r$, (III) $p=q=r/2$, and for the uncorrelated case (IV) $p=q$ and $r=0$ (this reduces to the random-bond Ising model on a square lattice). We find that as the correlated error strength $r$ increases, the thresholds decrease to $p_c(\text{I}) = 0.0925(25)$, $p_c(\text{II}) = 0.0725(25)$, $p_c(\text{III}) = 0.0475(25)$ from the uncorrelated threshold $p_c(\text{IV}) = 0.110(5)$. Near the Nishimori lines, longer MC runs are required to locate $p_c$ and $T_c$ accurately. The error in $p_c$ is determined within the present MC statistics (see Appendix~\ref{sec_monte_carlo_details}).

\begin{figure}
    \centering
    \includegraphics[width=0.6\textwidth]{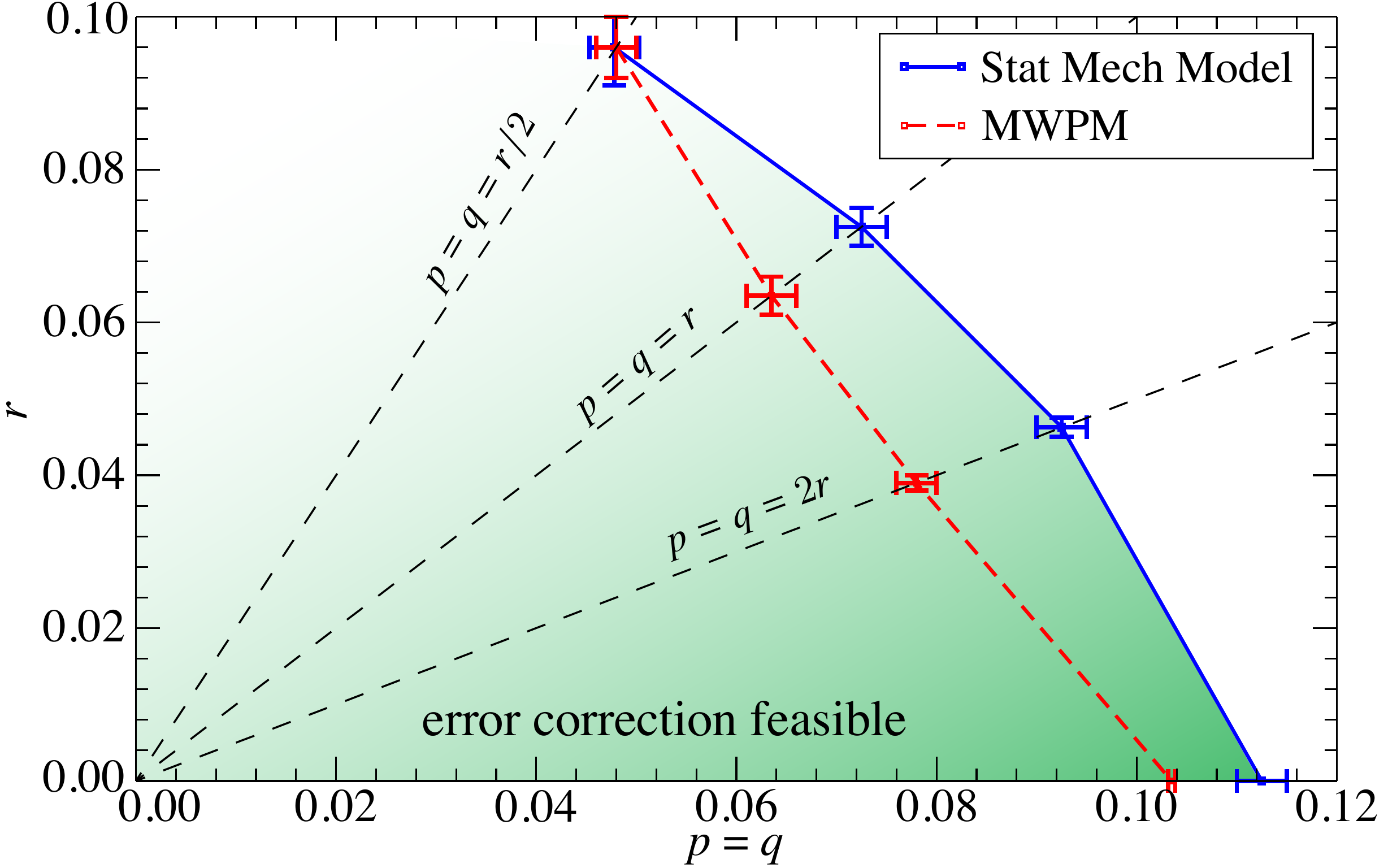}
        \caption{Threshold probabilities from the statistical mechanics model (blue points) and the MWPM decoder (red points). Lines are a guide to the eye. The region where the MWPM decoder finds correctability lies within the ordered region (green) of the statistical mechanics model. The latter region represents the fundamental regime, independent of specific decoding strategies, where correction via the repetition code is possible in principle.}
    \label{fig:rvsp}
\end{figure}

\subsection{Minimum weight matching decoder}

The thresholds extracted via MWPM provide lower bounds to the thresholds obtained through the statistical mechanics mapping since they correspond to a decoding decision that is not necessarily optimal: MWPM looks for the likeliest errors given a syndrome, which in the statistical mechanics mapping corresponds to minimizing the energy as opposed to the free energy~\cite{Dennis2002}. MWPM therefore disregards the degeneracy of errors, because the likeliest error need not necessarily be in the most likely class of equivalent errors. A discrepancy between MWPM and optimal (maximum-likelihood) decoding is known to exist already for phenomenological noise~\cite{Wang2003}. In order to make a fair comparison, we optimize the weight metric going into the MWPM algorithm, which uses the information available about the syndrome graph. A detailed derivation of the error weight metric used for decoding and information on the general MWPM strategy are provided in Appendix~\ref{sec_mwpm_decoder_details}.

\subsection{Discussion}
Figure~\ref{fig:rvsp} shows the thresholds from MC simulations (blue points, solid line) and from MWPM decoding (red points, dotted line). The region where MWPM successfully decodes lies within error bars inside the fundamental region of correctability determined with the statistical mechanics mapping. This demonstrates a finite interval between MWPM and a higher fundamental threshold that could be approached or achieved by improving the decoding strategy. To connect the phase diagram to experimental situations, let us discuss the relative strength of $r$ and its meaning. First, $p$ and $q$ both are typically dominated by similar error processes that for many qubit platforms are of comparable error rates, hence we set $p=q$ for simplicity, and additionally, we dial up the strength of $r$. The known case $r=0$ reduces to the phenomenological case without correlated errors, which would correspond to perfect two-qubit gates ($p_2=0$). Increasing $r$ to $r=p/2=q/2$ corresponds to $p_2=5p_1 +\mathcal{O}(p_1^2)$, which is roughly compatible with the two-qubit gate being an order of magnitude worse than single-qubit operations (error rate or infidelity) as observed in many experimental realizations  \cite{Huang2019,Chatterjee2021, Chen2021,Bruzewicz2019}. Considering $r=p=q$ corresponds to $p_2$ being dominant and all other error sources negligible ($p_\text{sp}=p_\text{id}=p_1=p_\text{m}=0$). The case $r=2p=2q$ goes beyond the depolarizing noise model that underlies Eq.~(\ref{eq:effective_error_rates}): not only is the two-qubit gate the only error source, it furthermore specifically produces errors leading to the $r$ type instead of $p$ and $q$, which 
would be described by an asymmetric depolarizing noise channel biased towards $r$. Finally, pure $r$ errors ($p=q=0$) correspond to a situation where syndrome information is perfectly trustable if interpreted correctly, akin to a repetition code with perfect measurements (with a threshold approaching 0.5 asymptotically). 
We thus see that the most experimentally relevant region $r\neq p \approx q$ is where we also find a clear separation between MWPM and the fundamental threshold. The separation seems to become narrower for larger contributions of~$r$, which is compatible with the expectation in the extreme case of only $r$ errors with an asymptotic threshold of 0.5. From a practical operational viewpoint, the estimated thresholds for ideal decoding and MWPM for $p=q=2r$ correspond to single-qubit errors $p_1\approx 0.02$ (ideal) vs.~$0.017$ (MWPM) and the two-qubit error rate $p_2\approx 0.087$ (ideal) vs.~$0.074$ (MWPM). For $p=q=r$ the threshold is $p_2\approx 0.14$ (ideal) vs.~$0.12$ (MWPM). 

Interestingly, in a recent experimental realization of the phase-flip code~\cite{Chen2021} circuit errors are reported to be well described by Pauli errors. Casting the experimentally obtained error rates into Eq.~(\ref{eq:effective_error_rates}), we find that they correspond to effective error rates $p=0.032$, $q=0.0285$ and $r=0.0035$, which is in excellent agreement with the observation that this experiment is thoroughly in the error-suppression regime.

\section{Outlook}

The techniques presented in this work can be extended to more complex complete QEC codes, including surface and color codes, and concatenated codes. We anticipate that as the code circuitry becomes more complex, it will be necessary to include more types of effective noise processes, which in turn will give rise to new statistical mechanics models with even richer interaction and disorder properties. It will also be interesting to extend the mapping technique to non-Clifford dynamics, and to temporally or spatially correlated circuit-level noise, to study the QEC potential of an even broader class of realistic quantum processors. 

\begin{acknowledgments}
We thank B. Lucini, C.~Chubb and S.~Flammia for useful discussions and B.M.~Terhal, E.~Fiorelli and P.~Pieri for their feedback on the manuscript. SK is supported by the National Research Foundation of Korea under grant NRF-2021R1A2C1092701 funded by the Korean government (MEST) and in part by NRF-2008-000458. MM acknowledges support by the ERC Starting Grant QNets Grant Number 804247, the EU H2020-FETFLAG-2018-03 under Grant Agreement number 820495, by US A.R.O. through Grant No. W911NF-21-1-0007, and by the Office of the Director of National Intelligence (ODNI), Intelligence Advanced Research Projects Activity (IARPA), via US ARO Grant number W911NF-16-1-0070. All statements of fact, opinions or conclusions contained herein are those of the authors and should not be construed as representing the official views or policies of ODNI, the IARPA, or the US Government. MR is supported by ERC grant EQEC No. 682726. MR would like to thank D.P. DiVincenzo for generously providing access to computing facilities.
\end{acknowledgments}

\bibliographystyle{mio_apsrev4-1}
\bibliography{abbr_biblio}

\appendix

\section{Effective error model from circuit level noise}\label{sec_derivation_circuit_model}
\begin{figure*}[h]
    \centering
    \includegraphics[width=0.7\columnwidth]{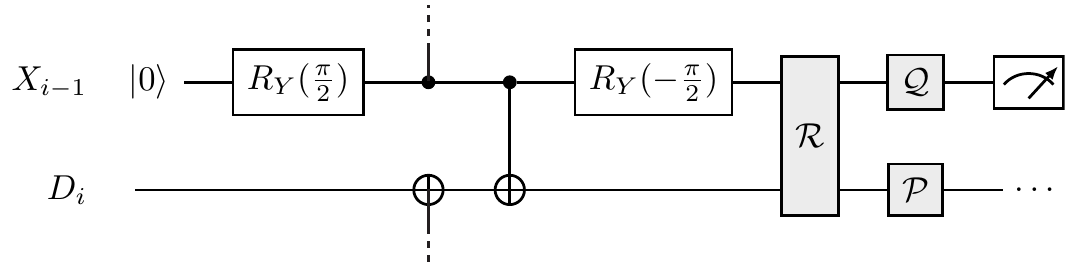}
    \caption{Cutout of one ancilla and one data qubit of the QEC circuit. All noise contributions of the circuit-level noise model (see Fig.~1(a) of the main text) can be merged into the three channels $\mathcal{P}$, $\mathcal{Q}$ and $\mathcal{R}$. The effective error rates are given in Eq.~\eqref{eq:effective_error_rates_supp}.}
    \label{fig:effective_noise_model}

  \end{figure*}

  Circuit-level noise in general, by definition, introduces noise channels at all locations inside the circuit. However for the given QEC code, it is possible to rewrite the noise model by exploiting the fact that many errors injected at different locations effectively lead to the same error and are thus equivalent, which allows us to arrive at a much simpler noise model (while keeping the mapping exact). Let us establish these equivalences as follows: on a data qubit, observe that $X$-errors do not affect the syndrome, hence for the purposes of decoding $X\sim \mathds{1}$ and thus also $Y\sim Z$. On ancilla qubits, the situation is similar, except that the intermediate rotations $R_Y(\pm\pi/2)$ flip the relevant error: at initialization and read-out, only $X$-errors are relevant, $Z$ does neither alter the initialization state nor the readout in the computational basis. Between the rotation gates $R_Y(\pm \pi/2)$, the roles are reversed and only $Z$ is relevant. This immediately lets us move all single-qubit channels to one side, since all noise channels commute (Pauli channels commute) and the Z-errors commute past the control of the CNOT. Similarly, the data qubit single noise channels can be collected into four consecutive single qubit noise channels. Turning to the two-qubit error channels, let us observe that by the equivalence of errors there are four possible distinct errors arising from the two-qubit depolarizing channel: $\mathds{1}\mathds{1}$,$\mathds{1}Z$, $Z\mathds{1}$ and $ZZ$ (omitting the tensor product symbol). There is a single new phenomenon encoded here: looking at the earlier CNOT noise channel, observe that $ZZ$ is equivalent to a simple data qubit error that would have preceded the CNOT and $\mathds{1}Z$ (data-ancilla) is equivalent to a simple measurement error,  $Z\mathds{1}$ generates a new syndrome phenomenon, which we call an $r$-type error. The effect of the second two-qubit noise channel is the same, just the roles of the terms being permuted. Assuming a depolarizing probability $p_2$ of the CNOT gate, defining the operators $P=\mathds{1}\otimes Z$, $Q=X\otimes \mathds{1}$ and $R=X\otimes Z = P\cdot Q$ and using hats for superoperator notation $\hat{M}(\rho):= M\rho M^\dag$ we thus have 
  \begin{equation}
    \mathcal{N}_{2qubit} = \left(1-\frac{12p_2}{15}\right)  \hat{\mathds{1}} + \frac{4p_2}{15}\left( \hat{P} +  \hat{Q} +  \hat{R}\right)
    \label{eq:two_qubit_channel}
  \end{equation}

  This channel is a mixture of three error events, which we would like to factorize into independent channels. Remarkably, we can make the ansatz of a product of three channels and solve for their error rates.
  \begin{equation}
    (\mathcal{N}_{2qubit})^2 =\left [(1-\lambda_p)\hat{\mathds{1}} + \lambda_p \hat{P}\right ]\cdot \left [(1-\lambda_q)\hat{\mathds{1}} + \lambda_q \hat{Q}\right ]\cdot \left [(1-\lambda_r)\hat{\mathds{1}} + \lambda_r \hat{R}\right ]
  \end{equation}
  Note that the square means applying the channel twice, which comes from the fact that we have two two-qubit noise channels acting subsequently, both of the form \eqref{eq:two_qubit_channel}. By symmetry of Eq.~\eqref{eq:two_qubit_channel}, the three rates must be equal, they turn out to be $\lambda_r = \lambda_p = \lambda_q = {8}p_2/15$. For the r-type error we are done, but for the effective single-qubit channels $\mathcal{P}$ and $\mathcal{Q}$, it remains to incorporate the four contributions from the single-qubit error channels. This is straightforward in the Pauli-transfer-matrix (PTM) representation~\cite{Greenbaum2015}, where we characterize a channel $\mathcal{E}$ by the matrix $R_{ij} =  \mathrm{tr}(P_i \mathcal{E}(P_j)) / d$ with $i \in \{0,1,2,3\}$. The strength of this representation is that here the action of a sequence of channels is simply given by matrix multiplication of their PTMs. The PTM of a bit-flip channel is
  \begin{equation}
      \hat{\mathcal{E}}=(1-\gamma)\hat{\mathds{1}}+\gamma \hat{X} \cong R_{\mathcal{E}} = \begin{pmatrix}  1 & 0 & 0 & 0\\ 0 & 1 & 0 & 0\\ 0 & 0 & 1-2\gamma & 0\\ 0 & 0 & 0 & 1-2\gamma\end{pmatrix}
  \end{equation}
  and furthermore it is evident that composing bit-flip channels results in a new bit-flip channel, such that we can read off the error rate from computing one of the non-trivial entries on the diagonal.  This results in $\mathcal{P} = (1-p)\hat{\mathds{1}} + p \hat{P}$, $\mathcal{Q} = (1-q)\hat{\mathds{1}} + q \hat{Q}$ and $\mathcal{R} = (1-r)\hat{\mathds{1}} + r \hat{R}$ with
  \begin{align}
    \begin{split}
    &p= \frac{1}{2} \left[1- \left(1-\frac{16p_2}{15}\right)\left(1-\frac{4p_\text{id}}{3}\right)^4 \right]\\
    &q= \frac{1}{2} \left[1- \left(1-\frac{16p_2}{15}\right)\left(1-\frac{4p_1}{3}\right)^2\left(1-\frac{4p_\text{sp}}{3}\right)\left(1-\frac{4p_\text{m}}{3}\right) \right]\\
    &r=\frac{8}{15}p_2.
    \end{split}
      \label{eq:effective_error_rates_supp}
  \end{align}

These are the three channels depicted in Fig.~\ref{fig:effective_noise_model}. Let us comment on the structure of these expressions: we recover the standard circuit-level noise by setting $p_2= p_\text{id} =p_\text{m} = p_1=p_\text{sp} = \lambda$. In that case, in leading order in $\lambda$ we recover $r/p = 1/6$, which can be understood also by simply counting the number of locations weighted by their error probability. While distinguishing between different error rates of different components in the circuit is straightforward, the factorization of the two-qubit noise channel into independent channels crucially relied on the symmetry between the different $P_i\otimes P_j$ terms. This suggests that biasing a particular two-qubit noise term would make the prefactor in Eq.~\eqref{eq:two_qubit_channel} non-symmetric and thus hinder the factorization of the channel. As a side note, we remark on a slightly non-intuitive feature in the conventional definition of the depolarizing channel: if we define it (as we do) as a random application of Paulis, complete depolarization (i.e. deterministically receiving the completely mixed state) corresponds to $\lambda=3/4$ (single qubit depolarizing channel) or $\lambda=15/16$ (two-qubit depolarizing channel), which shows that no term in the product inside Eq.~\eqref{eq:effective_error_rates_supp} becomes negative. Additionally, some works in the literature choose to not use a depolarizing noise for every location but rather put a bit-flip channel at the initialization and measurement locations. The two differ by a rescaling-factor of $2/3$, since only two of the three Pauli errors lead to errors at initialization and readout (e.g. in the computational basis a $Z$ acts trivially etc.).

\section{Derivation of the correlated disordered interacting classical spin model}\label{sec_derivation_stat_mech}
The technique of the statistical mechanical mapping is used to construct a classical statistical Hamiltonian with quenched disorder for describing the error model of a quantum code. The mapping relies on the identification of the partition function of the Hamiltonian with the probability $\Pr(\bar{E})$ of the class $\bar{E}=\{E'\}$ of errors $E'$ that are equivalent up to the action of space-time equivalences to a reference error $E$. Several ways have been introduced for deriving this mapping~\cite{Dennis2002, Chubb2019}. Here we will follow mainly the approach of Ref.~\cite{Chubb2019}.

Any error $E$ of the repetition code is composed of the fundamental errors $e_1, e_2, e_3$ introduced in the main text that in turn are generated by the three error processes (i) single-qubit phase flip error on data qubits with probability $p$, (ii) measurement error with probability $q$, and (iii) single-qubit phase flip errors happening on the target qubit between two adjacent CNOT gates that trigger a correlated occurrence of a measurement error on one of the neighboring ancilla qubits with probability $r$.

The probabilities of the errors $e_1$, $e_2$, $e_3$ are easily calculated and are given by 
\begin{equation}\label{supp_error_model}
\begin{aligned}
\pi_0 \equiv \Pr(e_0) & = (1-p)(1-q)(1-r) + p q r ,\\
\pi_1 \equiv \Pr(e_1) & = p (1-q)(1-r) + r q (1-p), \\
\pi_2 \equiv \Pr(e_2) & = q (1-p)(1-r) + r p (1-q), \\
\pi_3 \equiv \Pr(e_3) & = p q (1-r) + r (1-p)(1-q).
\end{aligned}
\end{equation}
where, for completeness, we added the probability of the trivial error $e_0$ that corresponds to the absence or the simultaneous presence of all of the three error processes. 

\begin{figure}
    \centering
    \includegraphics[width=0.8\textwidth]{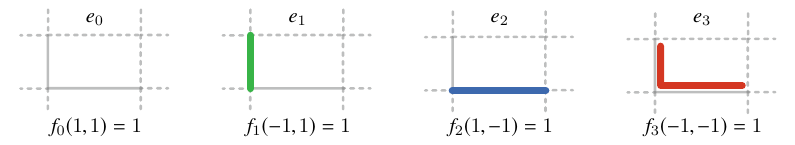}
    \caption{Fundamental errors $e_0$, $e_1$, $e_2$, $e_3$ parametrized by two binary variables $v, h \in \{-1,+1\}$ for the vertical and horizontal links and four Boolean functions $f_j(v,h) \in \{0, 1\}$. The trivial error $e_0$ corresponds to $(v,h) = (1, 1)$ and the function $f_0$ satisfies $f_0(1,1) = 1$. The error $e_1$ corresponds to $(v,h) = (-1, 1)$ and the function $f_1$ satisfies $f_1(-1,1) = 1$. The error $e_2$ corresponds to $(v,h) = (1, -1)$ and the function $f_2$ satisfies $f_2(1,-1) = 1$. The error $e_3$ corresponds to $(v,h) = (-1, -1)$ and the function $f_3$ satisfies $f_3(-1,-1) = 1$. On all the other cases not shown in the figure, the functions $f_j(v,h)$ are zero.\label{fig_f_functions}}
\end{figure}

In a section of the lattice in Fig.~1(c1) of the main text made only of a horizontal and a vertical link (shown also in Fig.~\ref{fig_f_functions}), the probability of a generic event $E$ composed by the errors $e_j$ with $j=0,1,2,3$ can be written as
\begin{equation}\label{supp_prob_E}
\Pr(E) = \pi_0^{f_0({v},{h})} \pi_1^{f_1({v},{h})} \pi_2^{f_2({v},{h})} \pi_3^{f_3({v},{h})}
\end{equation}
where $v, h\in\{\pm 1\}$ are binary variables that take the negative value $-1$ if the horizontal (vertical) link belongs to the error $E$. The Boolean functions $f_j\in\{0,1\}$ signal if $e_j$ belongs to the error $E$ and they satisfy $f_0(v=+1, h=+1) = f_1(v=-1, h=+1) = f_2(v=+1, h=-1) = f_3(v=-1, h=-1) =1$, otherwise they are zero. From this, they can be written as 
\begin{equation}\label{supp_f_binary_functions}
\begin{aligned}
f_0(v,h) = \frac{1}{4}(1 + h + v + v\, h), & \qquad
f_1(v,h) = \frac{1}{4}(1 + h - v - v\, h ),\\ 
f_2(v,h) = \frac{1}{4}(1 - h + v - v\, h ),& \qquad
f_3(v,h) = \frac{1}{4}(1 - h - v + v\, h ).
\end{aligned}
\end{equation}
By substituting the Eqs.~\eqref{supp_f_binary_functions} in Eq.~\eqref{supp_prob_E}, the probability $\Pr(E)$ can be written as 
\begin{equation}
\Pr(E) = \exp\left[-\beta H_E\right],
\end{equation} 
where $H_E = -J_0 - J_1 h - J_2 v - J_3 v\, h$ is a noise Hamiltonian \cite{Chubb2019} and the couplings $J_0, J_1, J_2, J_3$ satisfy
\begin{equation}\label{supp_Nishimori}
\begin{aligned}
\beta J_0 = \frac{1}{4}\log\left(\pi_0\pi_1\pi_2\pi_3\right)  , & \qquad \beta J_1 = \frac{1}{4}\log\frac{\pi_0\pi_1}{\pi_2\pi_3}, \\ 
\beta J_2 = \frac{1}{4}\log\frac{\pi_0\pi_2}{\pi_1\pi_3}   , & \qquad \beta J_3 = \frac{1}{4}\log\frac{\pi_0\pi_3}{\pi_1\pi_2}. 
\end{aligned}
\end{equation}
These are the Nishimori conditions connecting the Hamiltonian $H_E$ to the error model of Eq.~\eqref{supp_error_model}. The Hamiltonian  $H_E$ can be extended to the whole lattice by introducing binary variables  $h_\ell,v_\ell\in\{\pm 1\}$ associated to each link $\ell$ of the lattice. The variables $h_\ell$ ($v_\ell$) take the negative value $-1$ if the horizontal (vertical) link $\ell$ belongs to the error $E$. The Hamiltonian will then take the form
\begin{equation}\label{eqn_noise_hamiltonian_1}
H_E  = - \sum_\ell \left( J_0 + J_1 h_\ell + J_2 v_\ell + J_3 v_\ell\, h_\ell\right) .
\end{equation}

\begin{figure}
    \centering
    \includegraphics[width=0.25\textwidth]{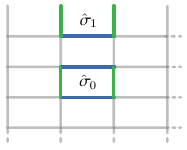}
    \caption{The equivalence $\hat{\sigma}_0$ in the bulk and $\hat{\sigma}_1$ at the boundary of the space-time lattice. The equivalence $\hat{\sigma}_0$  is given by the combined action of two  measurement  errors  on  the  ancilla qubits (horizontal blue links) preceded and followed by phase flip errors on the data qubit (vertical green links) adjacent to the ancilla qubits. The equivalence $\hat{\sigma}_1$ at the boundary is given by the combined action of one  measurement  error preceded and followed by phase flip errors on the data qubit. \label{fig_lattice_equivalences}}
\end{figure}

\begin{figure}
    \centering
    \includegraphics[width=0.8\textwidth]{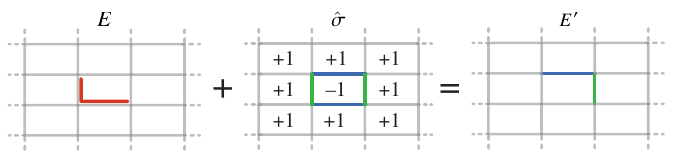}
    \caption{Action of the equivalence $\hat{\sigma}$ on an error $E=e_3$ (red L-shaped link). The equivalence $\hat{\sigma}$ (central panel) is given by two data-qubit errors (vertical green links) and two measurement errors (horizontal blue links). On the lattice the action of $\sigma$ is represented by a closed loop and it can be parametrized by a configuration of classical Ising spins $\{\sigma = \pm 1\}$ as explained in the text.\label{fig_code_equivalence}}
\end{figure}

The syndrome associated with the error $E$ can also be generated by other error chains $E'=E\hat{\sigma}$ that differ from $E$ by the action of the space-time equivalences $\hat{\sigma}$ that are trivial sequences of data-qubit errors and measurement errors that are not detected by the stabilizers.  Figure~\ref{fig_lattice_equivalences} shows the lattice with two highlighted equivalences: $\hat{\sigma}_0$ and $\hat{\sigma}_1$ belonging  to the bulk and the boundary of the lattice, respectively.

\begin{figure}
    \centering
    \includegraphics[width=0.6\textwidth]{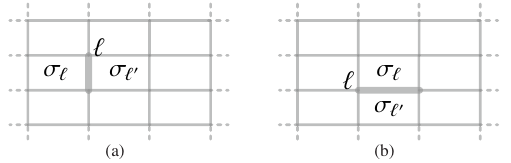}
    \caption{Parametrization of the equivalences and the error $E'$. For the equivalences: Given an Ising spin configuration $\{\sigma\}$ where each spin is associated to the plaquette of the lattice (see Fig.~\ref{fig_code_equivalence} central panel), (a) a data-qubit error occurring on the vertical link $\ell$ enters the equivalence $\hat{\sigma}$ if the product $\sigma_\ell \sigma_{\ell'} = -1 $ where the equivalences  $\sigma_\ell, \sigma_{\ell'}$ are on the left and on the right of $\ell$. (b) Measurement errors occurring on the horizontal link enters the equivalence $\hat{\sigma}$ if the product $\sigma_\ell \sigma_{\ell'} = -1$ where the equivalences  $\sigma_\ell, \sigma_{\ell'}$ are at the top and the bottom of $\ell$. For the error $E'$: given a reference error $E$ described by the binary variables $v_\ell$ and $h_\ell$ and the equivalences $\hat{\sigma}$, the error $E'$ equivalent to $E$ up to  $\hat{\sigma}$ is parametrized by $v'_\ell$ and $h'_\ell$ with (a) $v'_\ell = v_\ell \sigma_\ell \sigma_{\ell'}$ where the equivalences  $\sigma_\ell, \sigma_{\ell'}$ are on the left and on the right of $\ell$ and (b) for a horizontal link $\ell$, $h'_\ell$ is given by $h'_\ell = h_\ell \sigma_\ell \sigma_{\ell'}$ where the equivalences  $\sigma_\ell$ and $\sigma_{\ell'}$ are on top and at the bottom of $\ell$. \label{fig_error_e_prime}}
\end{figure}

We can associate classical Ising spins $\{\sigma\}$ to each of the equivalences such that if a spin $\sigma$ in this configuration is $-1$, we are effectively considering the error $E' = E \hat{\sigma}$ that differs from the reference error $E$ by the equivalence $\hat{\sigma}$ (for an example of an equivalence applied to an error $E$ see Fig.~\ref{fig_code_equivalence}). These Ising spins $\{\sigma\}$ can be associated to the plaquettes of the lattice in Fig.~1(c1) of the main text (see also Fig.~\ref{fig_code_equivalence}). The data-qubit errors (green vertical links in Fig.~\ref{fig_code_equivalence}) entering the equivalence $\hat{\sigma}$ are given by the vertical links $\ell$  of the lattice for which the product $\sigma_\ell \sigma_{\ell'} = -1 $ where the equivalences  $\sigma_\ell, \sigma_{\ell'}$ belong to the left and the right plaquettes w.r.t $\ell$ (Fig.~\ref{fig_error_e_prime}(a)). Similarly, the measurement errors (blue horizontal links in Fig.~\ref{fig_code_equivalence}) entering the equivalence $\hat{\sigma}$ are associated to the horizontal links $\ell$  of the lattice for which the product $\sigma_\ell \sigma_{\ell'} = -1$ where the equivalences  $\sigma_\ell, \sigma_{\ell'}$ belong to the top and the bottom plaquettes w.r.t. $\ell$ (Fig.~\ref{fig_error_e_prime}(b)).

The probability $\Pr(E')$ of the error $E'$ can also be written as $\Pr(E') = \exp\left[-\beta H_{E'}\right]$ where in the noise Hamiltonian  $H_{E'} = - \sum_\ell \left( J_0 + J_1 h'_\ell + J_2 v'_\ell + J_3 v'_\ell\, h'_\ell\right)$ the binary parameters $h'_\ell$ and $v'_\ell$ defining the error $E'$ are fixed by the parameters $h_\ell$ and $v_\ell$ of $E$ and by the equivalence $\sigma$ as follows:
\begin{itemize}
\item for a vertical link $\ell$, $v'_\ell = v_\ell \sigma_\ell \sigma_{\ell'}$ where the equivalences  $\sigma_\ell, \sigma_{\ell'}$ are on the left and on the right of $\ell$ (Fig.~\ref{fig_error_e_prime}(a));
\item for a horizontal link $\ell$, $h'_\ell$ is given by $h'_\ell = h_\ell \sigma_\ell \sigma_{\ell'}$ where the equivalences  $\sigma_\ell$ and $\sigma_{\ell'}$ are on top and at the bottom of $\ell$ (Fig.~\ref{fig_error_e_prime}(b)).
\end{itemize}

Substituting the variables $h'_\ell$ and $v'_\ell$ in the Hamiltonian $H_{E'}$ allows us to compute the probability $\Pr(\bar{E})$ of the class $\bar{E}=\{E'\}$ composed by errors $E'$ that are equivalent up to the action of equivalences to a reference error $E$ as
\begin{equation}\label{eqn_partition_function_code}
\Pr(\bar{E}) = \sum_{\{\hat{\sigma}\}} \Pr (E \hat{\sigma}) = \sum_{\{\sigma\}}  \exp\left[-\beta H_{E}(\sigma)\right],
\end{equation}
where
\begin{equation}\label{supp_eqn_noise_hamiltonian}
H_E(\sigma)  = -\sum_{\ell \updownarrow\ell'} J_1 h_\ell  \sigma_\ell \sigma_{\ell'} - \sum_{\ell \leftrightarrow \ell'} J_2 v_\ell  \sigma_\ell \sigma_{\ell'}  - \sum_{\ell \nwsearrow \ell' } J_3 h_\ell v_\ell  \sigma_\ell \sigma_{\ell'}
\end{equation}
is the Ising Hamiltonian with correlated disorder of the main text and we have neglected the unimportant additive constant $J_0$. The couplings $J_1$, $J_2$, $J_3$ and the inverse temperature $\beta$ are fixed from the error model describing $E$ by the Nishimori conditions previously obtained (see Eq.~\eqref{supp_Nishimori}). The type of interactions (ferromagnetic or antiferromagnetic) is given by the signs of $v_\ell$ and $h_\ell$ that are fixed by the reference error $E$. 
In the first sum of the r.h.s.~of Eq.~\eqref{supp_eqn_noise_hamiltonian}, $\sigma_\ell$ and $\sigma_{\ell'}$ are the spins horizontally adjacent to the link $\ell$ (e.g. spins $\sigma_1$ and $\sigma_2$ in Fig.~1(c1) of the main text); in the second sum, $\sigma_\ell$ and $\sigma_{\ell'}$ are the spins vertically adjacent to the link $\ell$ (e.g. spins $\sigma_3$ and $\sigma_4$ in Fig.~1(c1) of the main text); in the third sum, $\sigma_\ell$ and $\sigma_{\ell'}$ are the two spins  adjacent to the two arms of the L-shaped error $e_3$ (e.g. spins $\sigma_5$ and $\sigma_6$ in Fig.~1(c1) of the main text). Equation~\eqref{eqn_partition_function_code} shows that the probability $\Pr(\bar{E})$ of the class $\bar{E}$ can be written as the partition function $\mathcal{Z}_E = \sum_{\{\sigma\}}  \exp\left[-\beta H_{E}(\sigma)\right]$ of a statistical mechanical Hamiltonian $H_{E}(\sigma)$. The limiting behavior of the complementary probabilities $\text{Pr}(\bar{E}) $ and $ \text{Pr}(\overline{E Z_L})$ with increasing system size allows us to distinguish two regimes: below threshold, either of the probabilities converges to unity and applying a correction from the same error class will undo the error with probability approaching unity, above threshold all probabilities remain asymptotically finite~\cite{Bombin2012}.

\section{Monte-Carlo simulation study of the phase diagram}\label{sec_monte_carlo_details}
In this section we present the details on the numerical simulations of the Hamiltonian $H_E(\sigma)$. In particular, we describe how we choose the couplings $J_1, J_2, J_3$ of $H_E(\sigma)$ and the methods for obtaining the phase transition points.

\subsection{Couplings of the random bond-Ising model}\label{subsec_couplings}
The Hamiltonian of Eq.~\eqref{supp_eqn_noise_hamiltonian} that we analyse corresponds to a random-bond Ising model on a triangular lattice with couplings $J_1$, $J_2$ and $J_3$ (see Fig.~1(c2) of the main text). Relative ratio of these couplings are fixed by the Nishimori conditions (see Eq.~\eqref{supp_Nishimori})
while the type of interactions (ferromagnetic or antiferromagnetic) is assigned by drawing three random binary variables ${z_p, z_q, z_r} \in \{\pm 1\}$ with probability $p,q,r$  for each link of the triangular lattice. These variables take the negative sign if  one of fundamental processes (data-qubit error, measurement error, correlated error) is present in the reference error $E$. Therefore, they fix the signs of the variables $v_\ell$ and $h_\ell$ and thus of the couplings entering the Hamiltonian of Eq.~\eqref{supp_eqn_noise_hamiltonian} according to Table~\ref{table_couplings}.

\begin{table}\centering\begin{tabular}{|LLL|LL|ccc|}
\hline \hline
\multicolumn{3}{|c|}{Errors} & \multicolumn{2}{c|}{}   & \multicolumn{3}{c|}{Couplings}  \\ 
z_p  & z_q  & z_r   &  v_\ell & h_\ell & {$J_1 h_\ell$}  & {$J_2 v_\ell$}  & {$J_3 v_\ell h_\ell$} \\
\hline 
+1   & +1   & +1    &    +1   &   +1   &    F  & F  & F  \\
+1   & +1   & -1    &    -1   &   -1   &    AF  & AF  & F  \\
+1   & -1   & +1    &    +1   &   -1   &     AF  & F  & AF  \\
+1   & -1   & -1    &    -1   &   +1   &    F  & AF  & AF  \\
-1   & +1   & +1    &    -1   &   +1   &    F  & AF  & AF  \\
-1   & +1   & -1    &    +1   &   -1   &    AF  & F  & AF  \\
-1   & -1   & +1    &    -1   &   -1   &    AF  & AF  & F  \\
-1   & -1   & -1    &    +1   &   +1   &    F  & F  & F \\ 
\hline  \hline 
\end{tabular}\caption{The binary variables ${z_p, z_q, z_r}$, drawn randomly with probabilities $p,q,r$, represent the fundamental errors and fix the sign of the variables $v_\ell, h_\ell$ and thus of the couplings (ferromagnetic (F)  or antiferromagnetic (AF)) of the noise Hamiltonian of the random bond Ising model Eq.~\eqref{supp_eqn_noise_hamiltonian}.}\label{table_couplings}
\end{table}

\subsection{Finite-size scaling and transition points}
\begin{figure}
    \centering
    \includegraphics[width=0.85\textwidth]{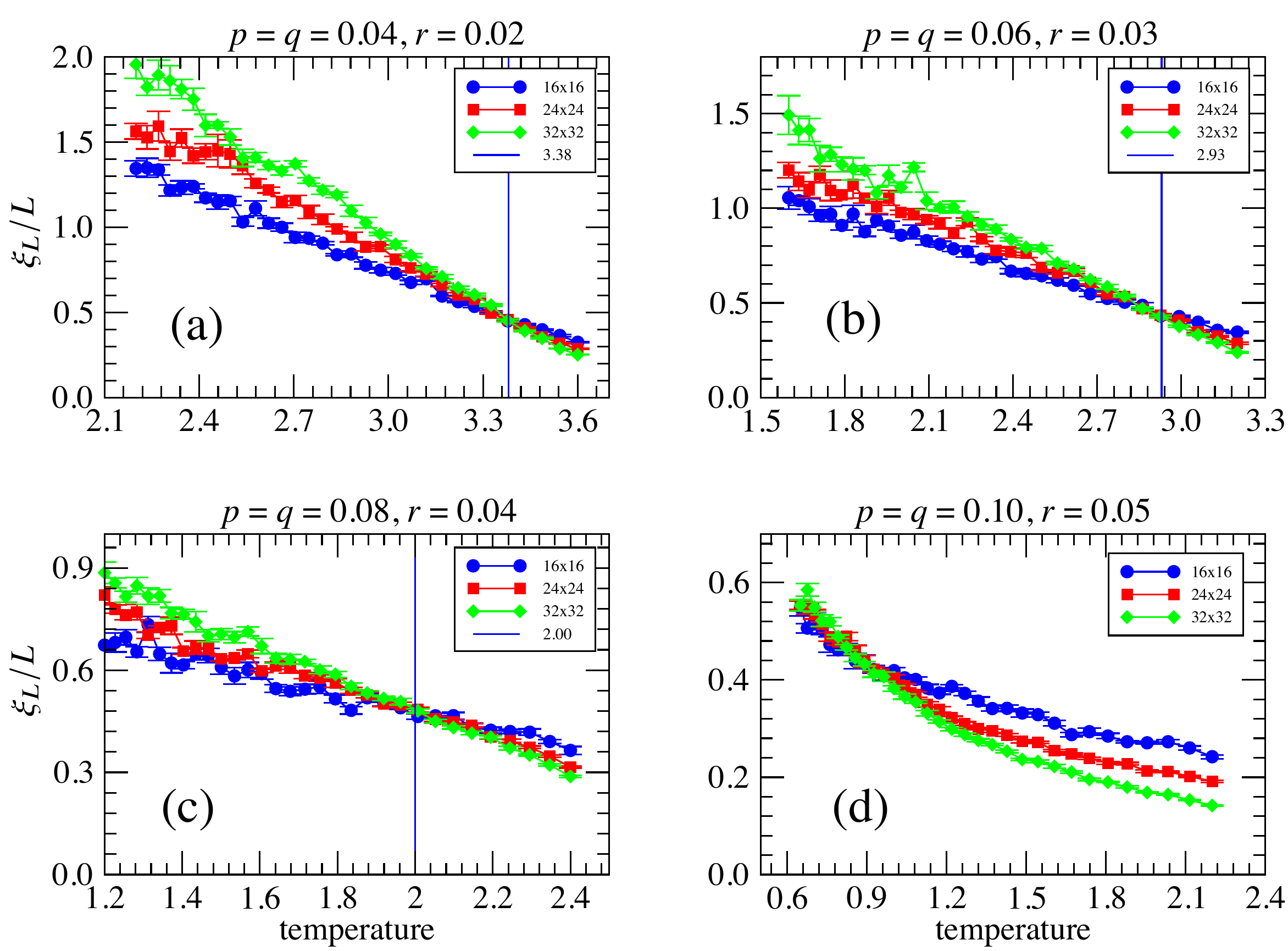}
    \caption{Critical temperatures from finite size scaling. The critical temperatures are found by locating the points where the lines representing the quantities $\xi_L/L$ cross each other for different systems sizes $L$. Examples of the occurrence of a phase transition are shown in (a) for $p = q = 0.04, r = 0.02$, in (b) for $p = q = 0.06, r = 0.03$, in (c) for $p = q = 0.08, r = 0.04$. If it is not possible to identify a temperature where the lines with the quantities $\xi_L/L$ cross, the transition does not occur. This is shown in panel (d) for $p = q = 0.10, r = 0.05$.}
    \label{fig:pq2r}
\end{figure}

Determining the code threshold requires in general the computation of  the scaling with the system size of the free energy cost of a domain wall as reported in Ref.~\cite{Dennis2002}. However since in~\cite{Chubb2019} the threshold of the quantum code has been proven to correspond to the phase transition point of the statistical mechanics model, in this work we locate the code threshold by looking at the critical point of the random Ising Hamiltonian. In the case of zero disorder the system is completely magnetized and a convenient order parameter is given by the total magnetization: $M = \sum_{\vec{x}} \sigma_{\vec{x}}$ where $\vec{x}$ denotes a site in the 2D lattice of linear size $L$ (here we consider lattices with $L = 16, 24$ and $32$). Instead of looking at the behavior of $M$ for different system sizes, we define the Fourier transform of the spin correlation function
$\langle \sigma_0 \sigma_{\vec{x}} \rangle$ as
\begin{equation}
  \hat{G}_L (\vec{k}) = \sum_{\vec{x}} \langle \sigma_0
  \sigma_{\vec{x}} \rangle_L \; e^{i \vec{k} \cdot \vec{x}}
\end{equation} 
and we extract the correlation length $\xi_L$ in the unit of lattice spacing from
\begin{equation}
  \xi_L = \frac{1}{2 \sin (q_{\rm min}/2)}
  \sqrt{\frac{\langle \hat{G}_L (0) \rangle_{\rm av}}{\langle
      \hat{G}_L (q_{\rm min}) \rangle_{\rm av}} - 1} 
\end{equation}
where $q_{\rm min} = (2\pi/L, 0)$. The brackets $\langle \cdots \rangle_{\rm av}$ denote averages over the quenched disorder distribution (in addition to the thermal average). Near the critical temperature $T_c$, the correlation length $\xi_L$ is expected to obey a scaling relation~\cite{Palassini1999} of the form $\xi_L/L \sim f(L^{1/\nu} (T - T_c))$, where $f$ is an unknown scaling function and $\nu$ is the critical exponent related to the correlation length. Therefore, at the critical temperature $T_c$, the quantity  $\xi_L/L $ becomes independent of the temperature and $T_c$ can be found by locating the temperature at which the lines representing $\xi_L/L$ for different $L$ intersect. Some examples of the comparisons of $\xi_L/L$ are given in Fig.~\ref{fig:pq2r} for the case $p = q = 2r$. In each figures, the vertical line indicates the temperature at which  $\xi_L/L$ cross for three different lattice sizes $L$. The figure without the vertical line shows the case in which there is no phase transition. 

For computing the spin correlation function and thus the correlation length we use the standard Metropolis algorithm~\cite{Metropolis1953} interspersed with parallel tempering between adjacent temperatures to allow efficient sampling of the low temperature phase~\cite{Swendsen1986,Earl2005}. Moreover, for a given quenched disorder distribution, the spin configurations are swapped after $N_\text{met} = 800$ Metropolis steps following the parallel tempering algorithm (except the data point for $p = q, r = 0$ with $p = 0.11$ (open blue circle in Fig.~2 of the main text) where $N_\text{met} = 8000$ Metropolis steps are performed). Metropolis steps/spin-swap combination are repeated 10000 times. The spin correlation function is measured during the Metropolis steps. Averages are taken over the Metropolis steps and then these averages serve as one of the jack-knife bins used to determine the statistical error. We use 250 different quenched disorder samples to compute the average of the correlation length over the disorder distribution. Table \ref{table_T_c} lists the details of the numerics: for every case shown in Fig.~2 of the main text, when the probability $p$ is fixed to the values reported in the second column of the table, we found the critical temperatures reported in the third column when the simulations are carried with a number of metropolis step $N_\text{met}$. 

\begin{table}\centering\begin{tabular}{|c
							 |S[table-format=2.4, round-precision=3, round-mode=places]
							 |S[table-format=2.3(1)]
							 |c|}
\hline \hline
case & $p$ & $T_c$ & $N_\text{met}$ \\
\hline
(I) $p = q = 2r$        & 0.001 & 3.86(2) & 800 \\
                        & 0.02  & 3.72(2) & 800 \\
                        & 0.04  & 3.38(2) & 800 \\
                        & 0.06  & 2.93(2) & 800 \\
                        & 0.08  & 2.00(2) & 800 \\
                        & 0.09  & 1.43(2) & 800 \\  
\hline
(II) $p = q = r $           & 0.0   & 3.64(2) & 800 \\
                        & 0.02  & 3.21(2) & 800 \\
                        & 0.04  & 2.74(2) & 800 \\
                        & 0.06  & 2.19(2) & 800 \\
                        & 0.07  & 1.67(2) & 800 \\
\hline
(III) $p = q = r/2$     & 0.001 & 3.38(2) & 800 \\
                        & 0.02  & 2.75(2) & 800 \\
                        & 0.03  & 2.44(2) & 800 \\
                        & 0.04  & 2.07(2) & 800 \\
                        & 0.045 & 1.80(2) & 800 \\
\hline
(IV) $p = q, r = 0$     & 0.0   & 2.266(2)& 800 \\
                        & 0.06  & 1.760(5)& 800 \\
                        & 0.08  & 1.565(5)& 800 \\
                        & 0.10  & 1.32(1) & 800 \\
                        & 0.105 & 1.16(1) & 800 \\
                        & 0.110  & 0.93(5) & 8000 \\
\hline \hline
\end{tabular}\caption{Critical temperatures $T_c$ for the different cases (I)-(IV) of quenched disorder from the Monte Carlo simulations. The columns report the case analyzed, the quenched disorder probability $p$, the corresponding critical temperature $T_c$ obtained by performing $N_\text{met}$ Metropolis steps.\label{table_T_c}}
\end{table}

\begin{table}\centering\begin{tabularx}{0.8\textwidth}{|c|>{\centering\arraybackslash}X|>{\centering\arraybackslash}X|}
\hline \hline
case & largest $p$  showing  a critical temperature & smallest $p$  not showing a critical temperature \\
\hline
(I) $p = q = 2 r$ & 0.090 & 0.095 \\
(II) $p = q = r $ & 0.070 & 0.075 \\
(III) $p = q =  r/2$ & 0.045 & 0.050 \\
(IV) $p = q, r = 0$ & 0.110 & 0.115 \\
\hline \hline
\end{tabularx}\caption{Lower and upper bounds of the probability $p$ between which the system exhibits a  critical temperature for the cases (I)-(IV). The threshold probabilities and the associated errors shown in Fig.~3 of the main text for the statistical mechanics model are estimated by the mean and the deviation of these bounds for $p$.} \label{table_threshold}
\end{table}

\begin{figure}
    \centering
    \includegraphics[width=0.85\textwidth]{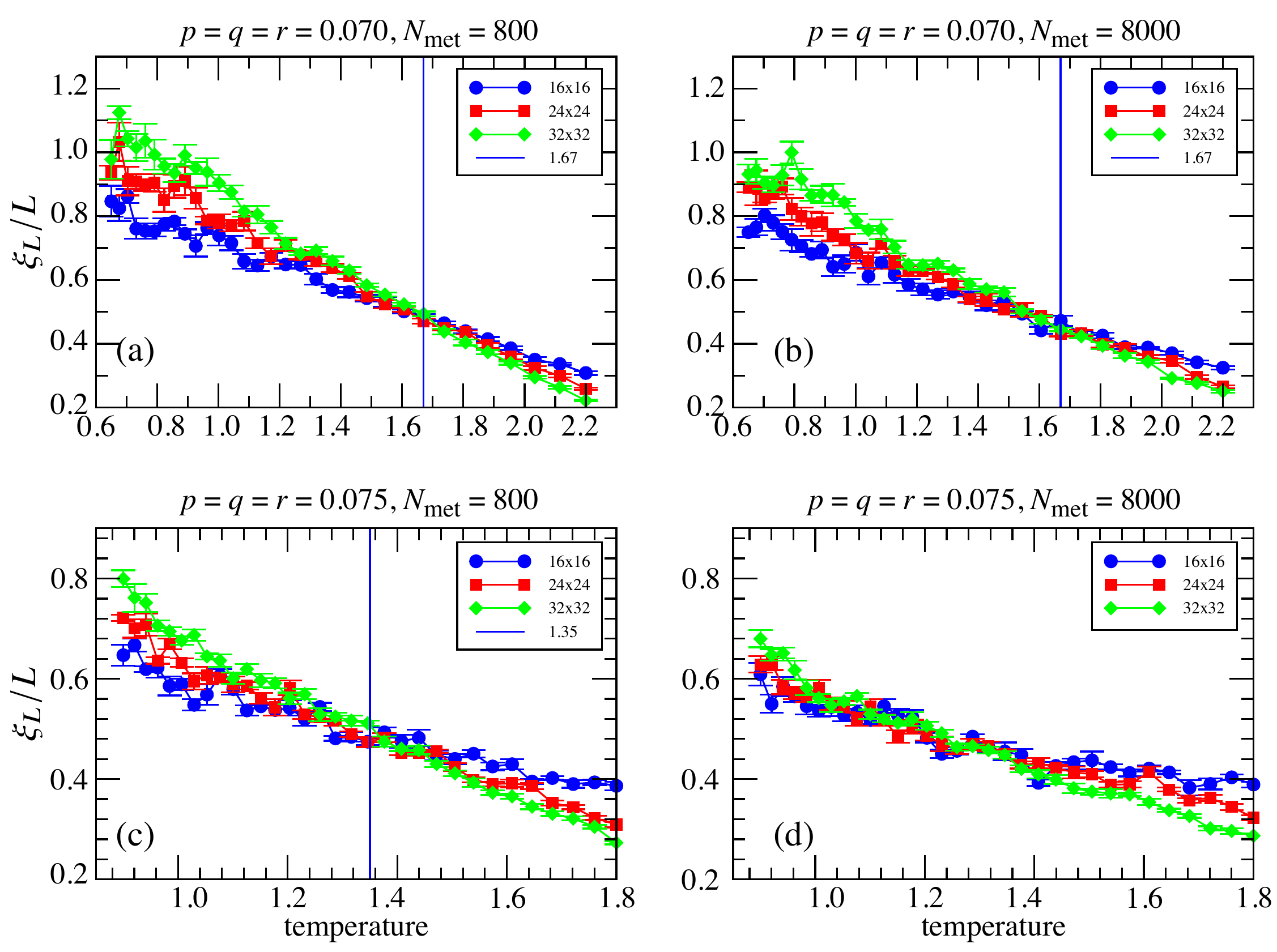}
    \caption{Effect of increasing the Metropolis steps. The quantity $\xi_L/L$ computed with (a) 800 Metropolis steps and (b) 8000 Metropolis steps is shown for $p = q = r = 0.07$ before the Nishimori line is crossed. In this case a transition is identified at a temperature $T_c = 1.67$. When instead $p = q = r = 0.075$ after the Nishimori line is crossed, a transition at $T_c = 1.35$ seems to appear when $\xi_L/L$ is computed with 800 Metropolis steps (panel (c)). However, when we compute $\xi_L/L$ with 8000 Metropolis steps (d) the lines do not cross any more and overlap for the temperature less than 1.35. This implies that a transition does not occur when $p = q = r = 0.075$.}
    \label{fig:pqrNishi}
\end{figure}

For assessing whether the random bond Ising model has thermalized we perform additional checks near the quenched probabilities at which the thermal transition crosses the Nishimori lines (Fig.~2 of the main text). In Fig.~\ref{fig:pqrNishi}, we compare the effects of increasing the Metropolis steps between parallel tempering spin-swap steps. 
Comparing the two top panels of Fig.~\ref{fig:pqrNishi}, we observe that there is no noticeable difference in the crossing of the lines $\xi_L/L$ when the number of Metropolis steps is increased from 800 (panel (a)) to 8000 (panel (b)) for the probability $p = q = r = 0.070$. On the other hand, the comparison of the two bottom panels in Fig.~\ref{fig:pqrNishi} ($p = q = r = 0.075$) shows that the crossing of $\xi_L/L$ for the three different lattice sizes, which seems to suggest a phase transition in panel (c) for 800 Metropolis steps, almost disappears for 8000 Metropolis steps (panel (d)) and the lines of $\xi_L/L$ overlap for all the temperatures $T\lesssim 1.35$. Therefore we conclude that there is no transition at the quenched disorder probability $p = q = r = 0.075$ and the threshold probability should be larger than $0.070$ but smaller than $0.075$. 

In Table~\ref{table_threshold}, we list the largest $p$'s at which MC simulations (run for $N_\text{met} = 8000$) show a phase transition and the smallest $p$'s at which MC simulations do not show a phase transition for the cases (I)-(IV). Figure~\ref{fig:rvsp} in the main text is produced with these data. The central value of these two $p$'s is taken as the estimated threshold probability and a half of the difference is given as the error estimate. A more precise determination of the threshold probability will require even longer Metropolis updates between parallel tempering steps.

\begin{figure}
    \centering
    \includegraphics[width=0.4\textwidth]{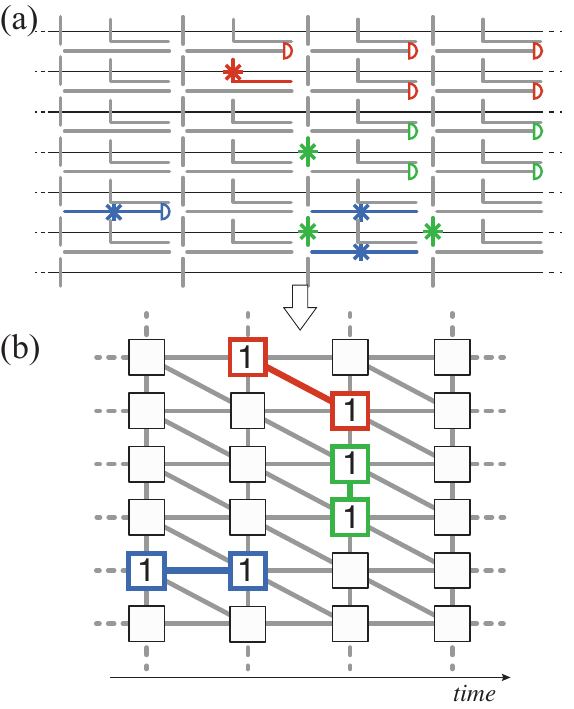}
    \caption{(a) Example of an error graph generated by physical errors. Positions where qubit, measurement and correlated data phase-flip and measurement errors can happen are represented by vertical, horizontal and L-shaped edges, respectively. Colored links represent occured error events while colored semicircles represent the ancilla qubits  triggered by an error event. From this graph we derive syndrome volume which is the input for the MWPM decoder. (b) Syndrome volume graph derived by the error graph in (a). The boxes labeled with 1s represent defects, i.e. ancilla qubits where the measurement outcome differs from the previous round of measurements. The colored diagonal red, vertical green and horizontal edges represent a possible matching of the defects.}
    \label{fig_decoder}
\end{figure}

\begin{figure}
\centering
\includegraphics[width=0.4\textwidth]{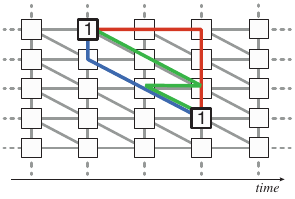}
\caption{Simplification in computing the distance of two defects (boxes labeled with 1s) in the triangular lattice. Depending on the values of the weights $w_p, w_q, w_r$ the shortest path between any two vertices can always be found by considering error patterns consisting of two types of errors only. If $w_r> w_p + w_q$ ($pq$ case) the shortest path is the regular Manhattan distance of the square lattice, counting the number of $p$ and $q$ edges between the two defects (red path). If $w_r + w_p < w_q$ ($pr$ case) the shortest path is composed by only $p$ and $r$ edges (blue path). Similarly, if $w_r + w_p > w_q$ ($qr$ case) the shortest path is composed by only $q$ and $r$ (green path).}\label{fig_distances}
\end{figure}

\section{Minimum-weight-perfect-matching decoding}\label{sec_mwpm_decoder_details}
To motivate this decoding strategy, observe that the probability of some error pattern $E$ consisting of $n_i$ errors of the $i-$th type, where $i=p,q \mathrm{\ or\ } r$) is proportional to
\begin{equation}
  p(E) \propto \left(\frac{p}{1-p}\right)^{n_p(E)} \left(\frac{q}{1-q}\right)^{n_q(E)} \left(\frac{r}{1-r}\right)^{n_r(E)}. \label{error_probability_mwpm}
  \end{equation}
  Given some observed collection of stabilizer measurement outcomes (the syndrome), repairing that syndrome amounts to finding a configuration of errors where every violated syndrome (a defect) is the endpoint of one error string, an assignment known as a (perfect) matching in graph theory (Fig.~\ref{fig_decoder}). Now Eq.~(\ref{error_probability_mwpm}) implies that finding the most likely among all these is equivalent to finding an error configuration with the lowest weight $w_{\mathrm{tot}} = -\log(p(E)) =  {n_p}{w_p} + {n_q}{w_q} + {n_r}{w_r}$, hence \emph{minimum weight} perfect matching. By creating a graph with all defects as vertices and the weight between every pair of nodes given by the number of error locations times error type weight ($w_p = -\log\frac{p}{1-p}$ etc.) between the respective stabilizers, this is what is solved by the above mentioned Blossom algorithm (implemented e.g. in \cite{networkx}). As a side remark, we note that the QEC code at hand does have a boundary, i.e. error strings extending to a boundary qubit only create a single defect, which is not immediately amenable to the above
  but can be incorporated by creating a copy of the matching graph with weights set to zero and then putting an edge between every defect and its virtual partner with the corresponding weight of matching to the code boundary \cite{Wang2010}. Due to the addition of $r$-type edges the calculation of weights going into the matching problem has to be adapted as well. While in general we would have to compute the shortest path on a triangular lattice with the three types of edges being weighted respectively, which can be done by a Dijkstra-type algorithm, we can make a vast simplification by observing that the shortest path between any two vertices in our setting can always be found by considering error patterns consisting of two types of errors only. To see this, observe that any single edge in a certain path can be replaced by the other two edges in the triangle, which are hence of the other two types, e.g. a particular $r$-edge can be circumvented by moving across the neighboring $p$- and $q$-edge. Furthermore all vertices are connected to at least one edge of each type and the weights are globally the same (up to the edge type). Let us assume that $w_r < w_p + w_q$ (if this was not the case the metric would not change compared to the case without $r$-edges). Now the two possible cases are that either $w_r+w_p < w_q$ or $w_r+w_p>w_q$. In the former case, we replace all $q$-moves by the same amount of $r$- and $p$- moves, in the latter case we do the opposite and replace all $p$-moves by $q$- and $r$-moves, which shows that given a path, we can always find a path of lower weight by eliminating one of the edge types. In the borderline case $w_r + w_p = w_q$ both replacements are admissible, such that the statement holds that we can eliminate one edge type (see Fig.~\ref{fig_distances}). This shows that a minimum weight path between any two vertices necessarily lies on one of the three sublattices and the path weight on a sublattice is given by the Manhattan distance on the respective sublattice:
\begin{equation}
  wt(s_1,s_2) = \min \begin{cases} ||s_1-s_2||_1^{pq}\\ ||s_1-s_2||_1^{pr}\\ ||s_1-s_2||_1^{qr} \end{cases}.
\end{equation}

\begin{figure}[h]
\centering
\includegraphics[width=0.4\textwidth]{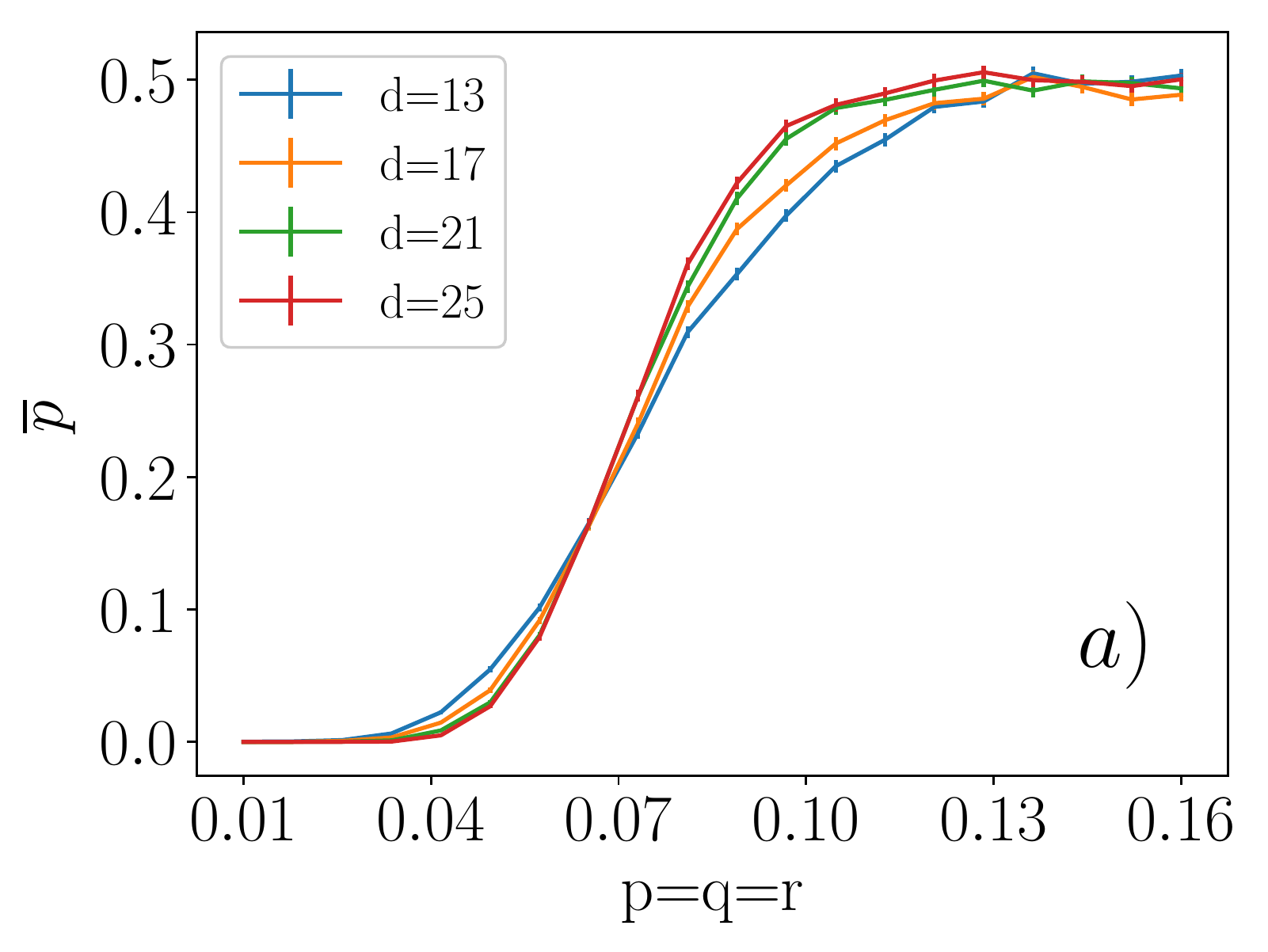}%
\qquad%
\includegraphics[width=0.4\textwidth]{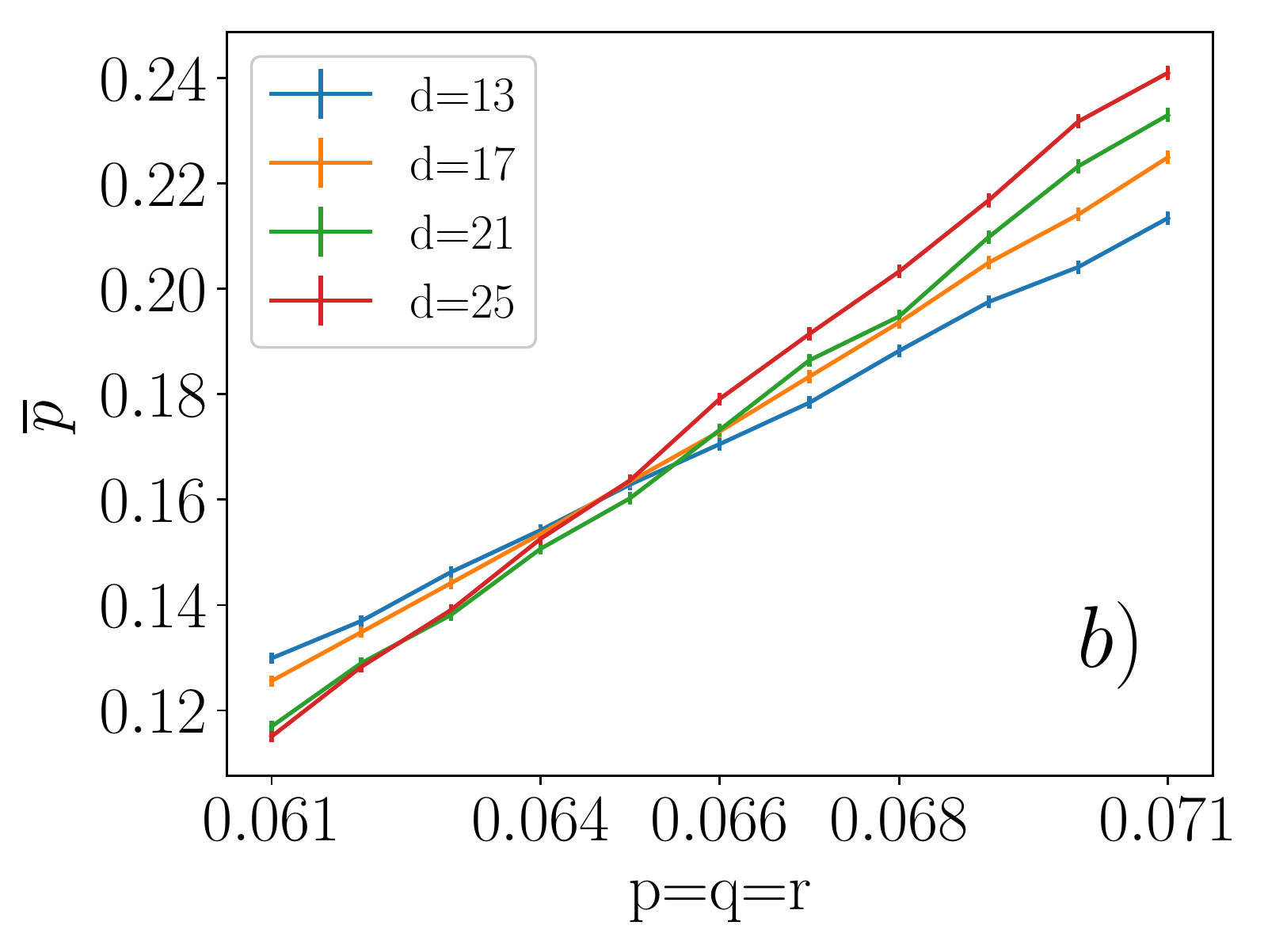}
\includegraphics[width=0.4\textwidth]{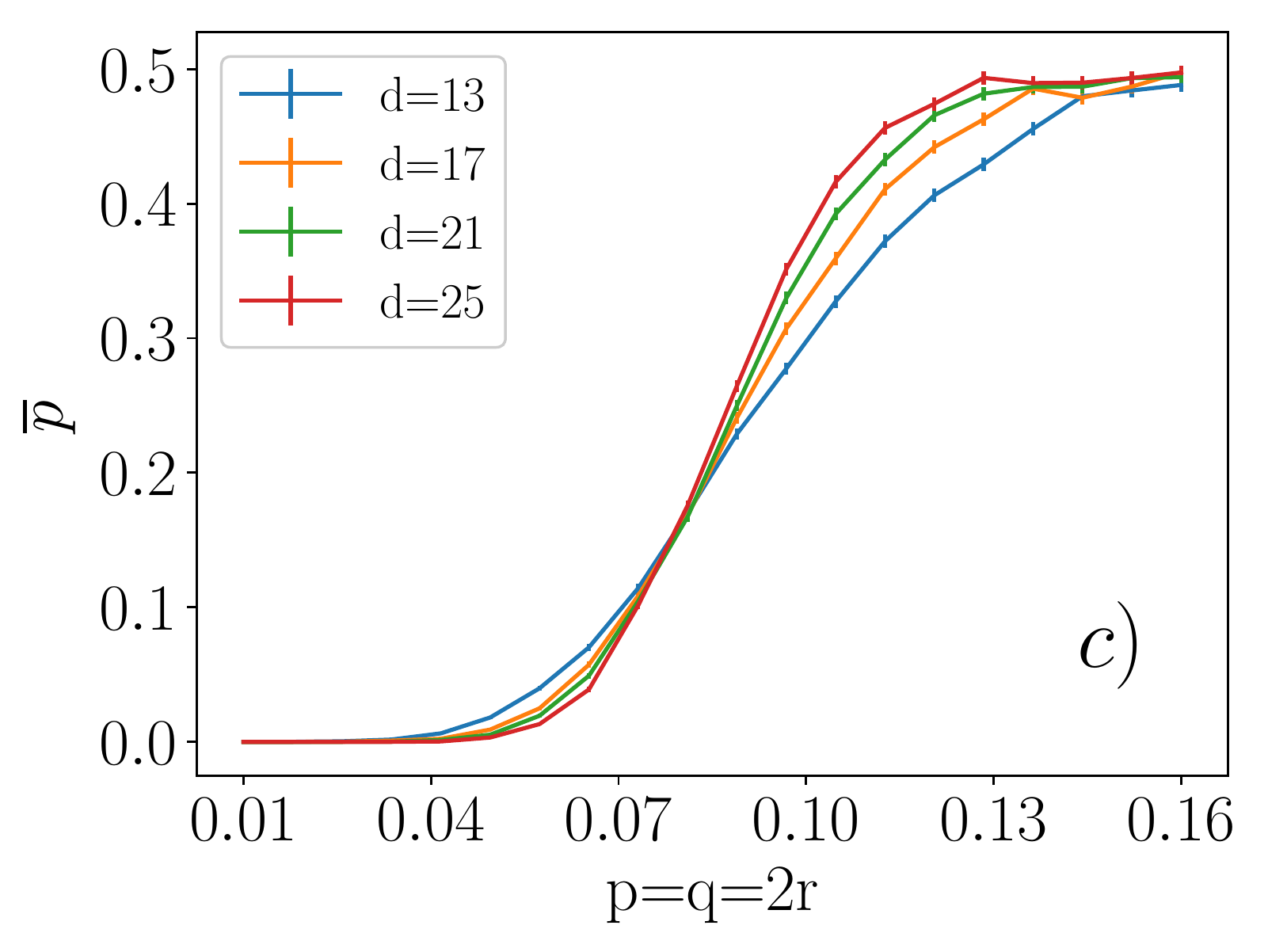}%
\qquad%
\includegraphics[width=0.4\textwidth]{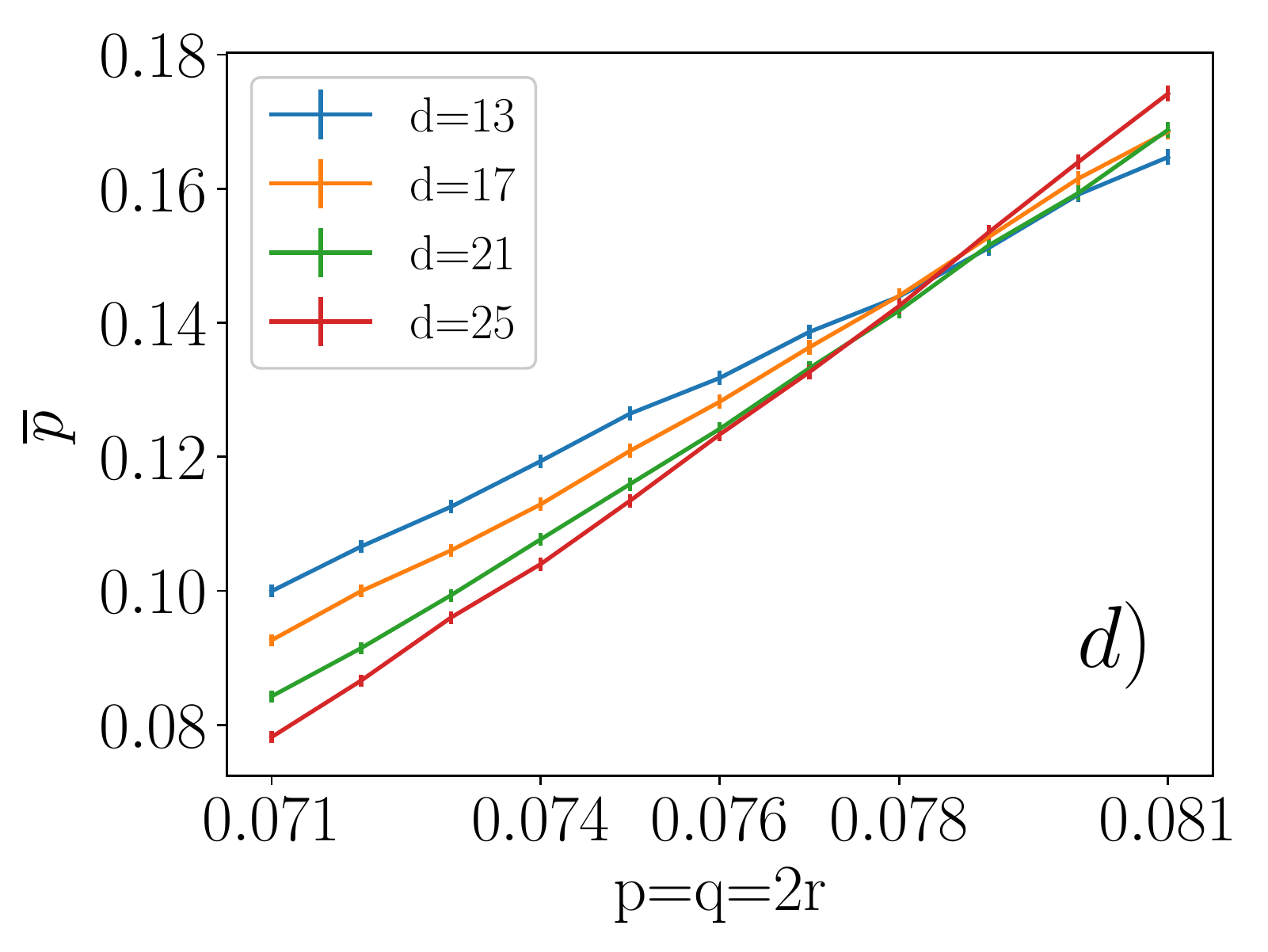}
\includegraphics[width=0.4\textwidth]{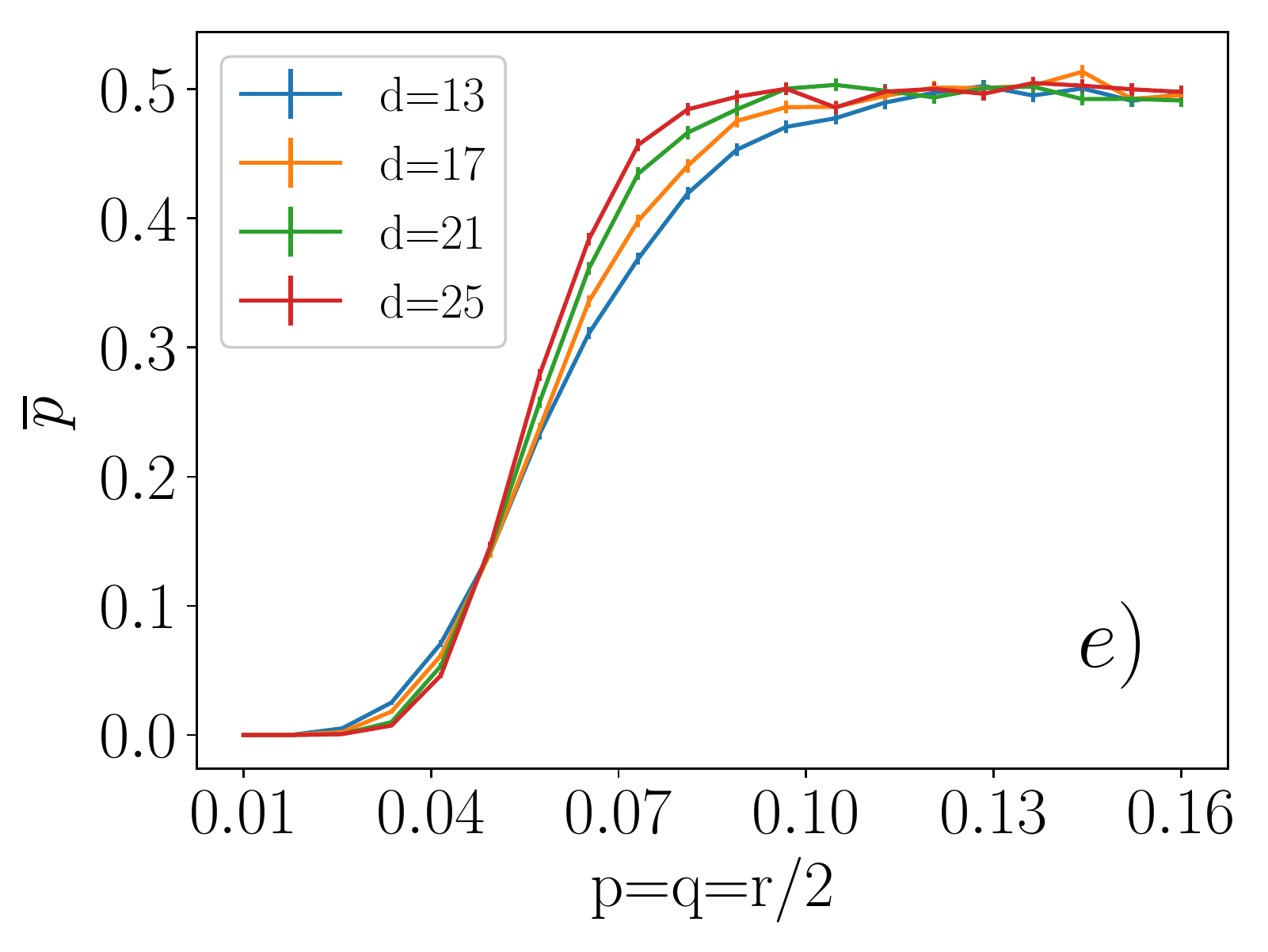}%
\qquad%
\includegraphics[width=0.4\textwidth]{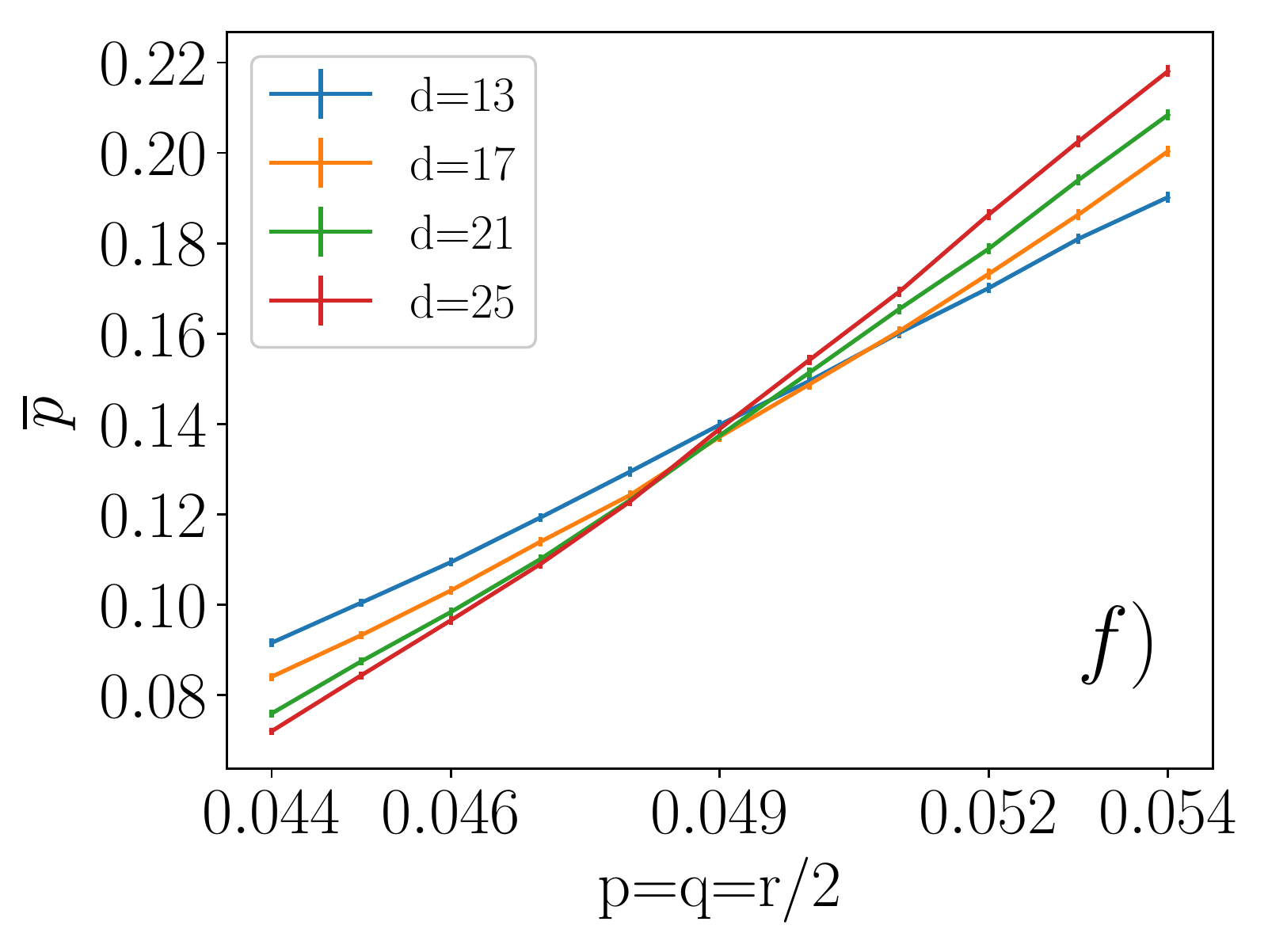}
\includegraphics[width=0.4\textwidth]{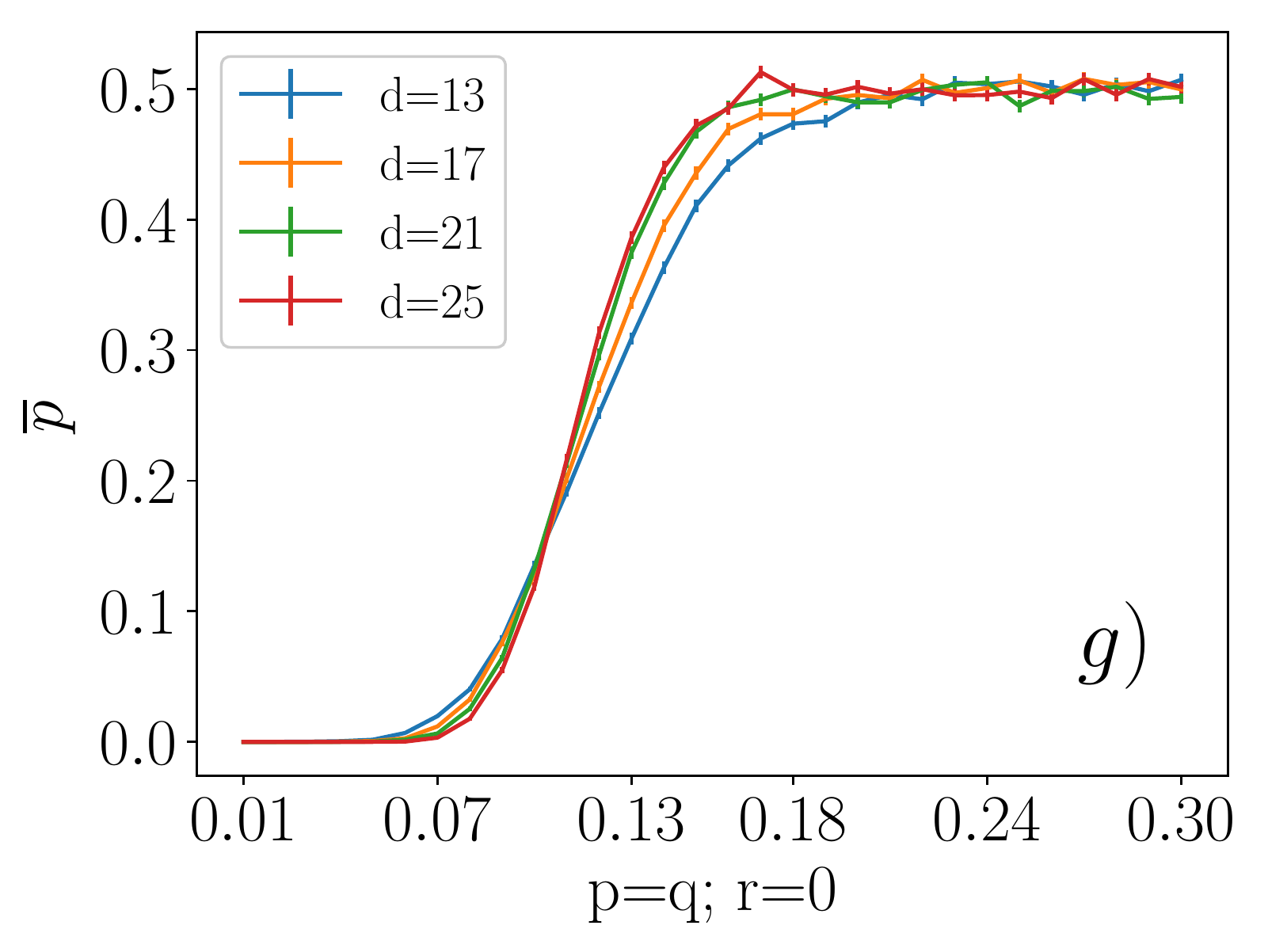}%
\qquad%
\includegraphics[width=0.4\textwidth]{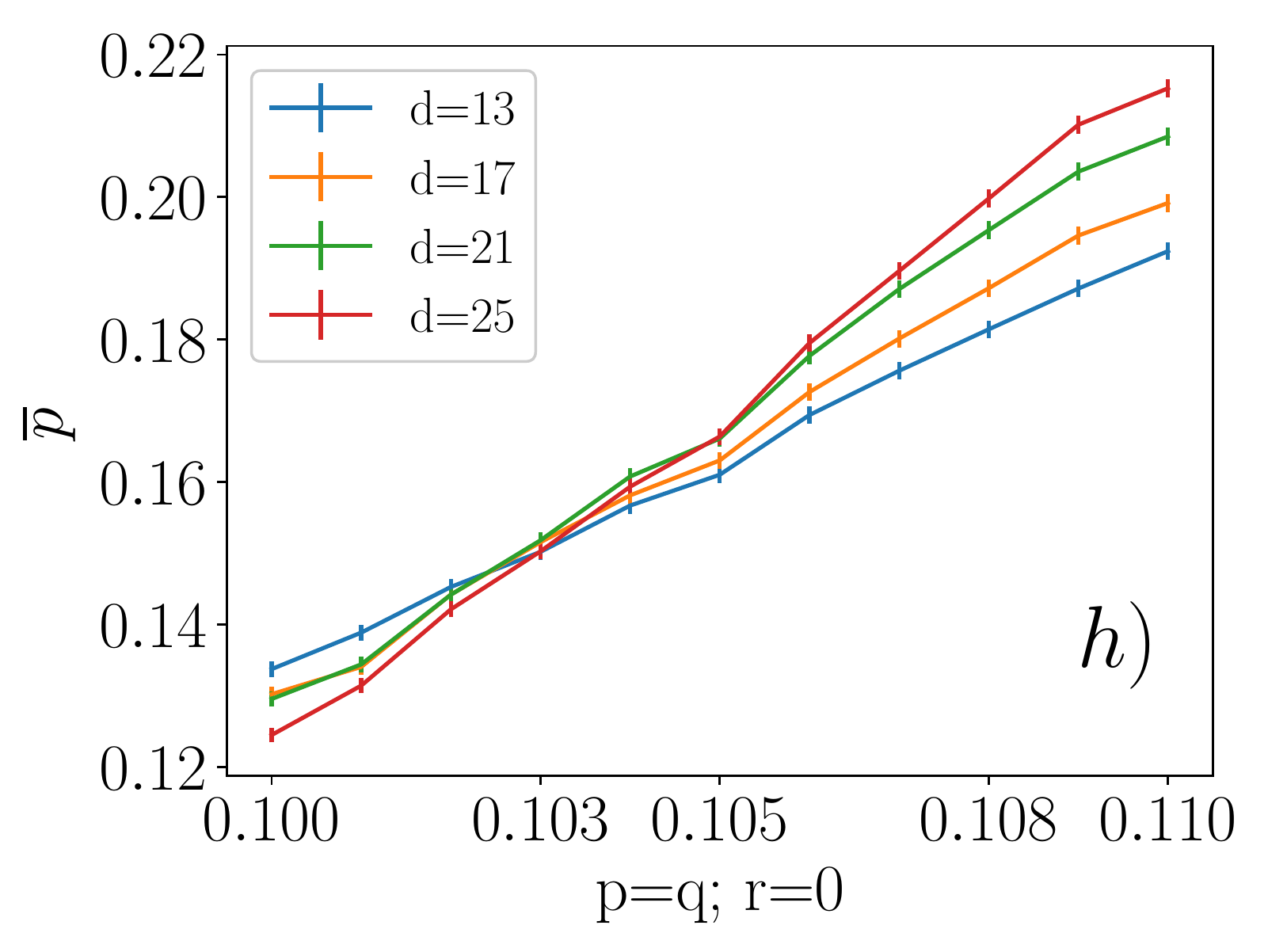}
 \caption{The logical error rate obtained with the minimum weight perfect matching decoder on the triangular syndrome lattice. Left column is a broad sweep, right column is a closeup around the transition point. We plot the average logical error rate as a function of the input error rates $p$, $q$, and $r$. Shown are the three cases $p=q=r$ in a)+b), $p=q=2r$ in c)+d), $p=q=r/2$ in e)+f) and $p=q, r=0$ in g)+h) . On the left for the broader scan, we use $10^4$ samples, data points in the right colum are averaged over $10^5$ samples. In each case, we observe a transition from error suppression to error enhancement with increasing error rate. This transition is signified by the behavior of the logical error rate when increasing the code size (the distance  d): for small p, increasing the distance leads to a suppression of the logical error rate, whereas for error rates beyond an inflection point increasing the code size instead leads to an increased logical error rate. We estimate the threshold from the region where we can not distinguish the logical error rates within errorbars.}
    \label{fig:MWPM_results}
\end{figure}

Here the two superscript labels indicate the sublattice of the triangular lattice, that consists of the two types of edges. In the $pq$ case this is just the regular Manhattan distance, counting the number of $p$ and $q$ edges between the two vertices at hand. In case of $pr$ and $qr$ this is the Manhattan distance on a skewed square lattice, but by simply rotating the basis vectors we can still use the Euclidean position vectors to compute the Manhattan distance on the respective sublattices. The weight function assigning the weight between any two defects then simply takes the minimum over the three weighted Manhattan distances. Having assigned the edge weights, we can then solve the minimum weight matching problem, which amounts to finding the most likely collection of errors explaining the observed syndrome in a trial run, from which we can therefore deduce the recovery operation we should apply to the code (or in which way we should update the ``Pauli frame''). The figure of merit we are after is the logical error rate, the probability that the decoding strategy fails. In order to cleanly decide in our simulation whether a logical error happened or did not happen, we initialize a perfect codeword, we then simulate $d$ subsequent QEC-cycles, each consisting of injecting phase-flip errors with probability $p$ on each data qubit followed by a noisy syndrome extraction (the syndrome bit being flipped with probability $q$). We furthermore inject the third error type needed, namely a correlated flip of the data-qubit accompanied by a flip of one of the syndrome measurement outcomes adjacent to it (depending on the order of the two-qubit gates this deterministically happens on the syndrome bit sitting either to the left or to the right of the data qubit). This type of error is injected with probability $r$. In order to make sure that we can declare the binary outcome of the trial run logical error/no logical error, we let the final measurement be perfect, i.e. in the final round the syndrome bits never flip (with $q$ or $r$). The final measurement being perfect ensures that matching up all defects will with certainty put the data qubits back into a code state. The simulation results are shown in Fig.~\ref{fig:MWPM_results}.

\end{document}